\documentclass[aps,pre,showpacs,amsmath,amssymb,amsfonts,superscriptaddress,twocolumn,lengthcheck]{revtex4-1}

\usepackage{graphicx}
\usepackage{subfigure}
\usepackage{verbatim}
\usepackage{dcolumn}
\usepackage{bm}
\usepackage{epsf}
\usepackage{color}
\usepackage[colorlinks=true,citecolor=blue,linkcolor=blue]{hyperref}

\newcommand{\bra}[1]{\left\langle #1\right|}
\newcommand{\ket}[1]{\left|#1\right\rangle}

\newcommand{\ptr}[2]{\mathrm{tr_{#1}}\left\{#2\right\}}

\newcommand{\la}{\left\langle}
\newcommand{\ra}{\right\rangle}

\newcommand{\td}{\mathrm{d}}

\newcommand{\e}[1]{\exp{\left(#1\right)}}

\newcommand{\bla}{bla\\bla\\bla\\bla\\bla}

\newcommand{\PRA}{Phys. Rev. A }

\newcommand{\PRE}{Phys. Rev. E }
\newcommand{\PRL}{Phys. Rev. Lett. }

\newcommand{\NJP}{New. J. Phys. }

\newcommand{\mc}[1]{\mathcal{#1}}

\begin{document}

\title{Interference of Identical Particles and the Quantum Work Distribution}

\author{Zongping Gong}
\affiliation{School of Physics, Peking University, Beijing 100871, China}

\author{Sebastian Deffner}
\affiliation{Department of Chemistry and Biochemistry and Institute for Physical Science and Technology,\\
University of Maryland, College Park, MD 20742 USA}
\affiliation{Theoretical Division and Center for Nonlinear Studies, Los Alamos National Laboratory, Los Alamos, NM 87545, USA}

\author{H. T. Quan}
\email{htquan@pku.edu.cn}
\affiliation{School of Physics, Peking University, Beijing 100871, China}
\affiliation{Collaborative Innovation Center of Quantum Matter, Beijing 100871, China}

\date{\today}

\begin{abstract}
Quantum mechanical particles in a confining potential interfere with each other while undergoing thermodynamic processes far from thermal equilibrium. By evaluating the corresponding transition probabilities between many-particle eigenstates we obtain the quantum work distribution function, for identical Bosons and Fermions, which we compare with the case of distinguishable particles. We find that the quantum work distributions for Bosons and Fermions significantly differ at low temperatures, while, as expected, at high temperatures the work distributions converge to the classical expression. These findings are illustrated with two analytically solvable examples, namely the time-dependent infinite square well and the parametric harmonic oscillator.
\end{abstract}


\maketitle

\newpage

\section{Introduction}

In the past two decades, nonequilibrium work relations \cite{Jarzynski2011}, including the Jarzynski Equality \cite{Jarzynski1997a,Jarzynski1997b} and the Crooks Fluctuation Theorem \cite{Crooks1998,Crooks1999} have attracted a lot of attention. These two nonequilibrium work theorems together with other exact relations concerning entropy production in arbitrary far-from-equilibrium processes, collectively known as Fluctuation Theorems \cite{Evans1993,Evans1994,Gallavotti1995a,Gallavotti1995b,Kurchan1998,Lebowitz1999,Hummer2001,Hatano2001,Seifert2005,Seifert2012}, have shed new light on our understanding of nonequilibrium thermodynamics beyond the close-to-equilibrium regime. The validity of the classical version of these relations has been tested experimentally in various systems \cite{Wang2002,Liphardt2002,Collin2005,Bustamante2005,Douarche2005a,Douarche2005b,Blickle2006,Douarche2006,Saira2012,Ciliberto2013}. In recent years, the quantum version \cite{Tasaki2000,Kurchan2000,Mukamel2003,Talkner2007} of these relations has also been proposed and studied extensively \cite{Huber2008,Esposito2009,Campisi2009,Campisi2011,Kihwan2014}. In the quantum regime, the so-called two-time energy measurement approach has proven to be effective. Within this approach quantum work performed by a thermally isolated system is determined by two projective energy measurements. On the other hand, the analysis of the characteristic function, i.e., the Fourier transform of the work density \cite{Talkner2007}, has opened new, alternative avenues to experimentally test quantum work theorems \cite{Dorner2013,Mazzola2013,Serra2014}.

Previous studies of quantum work relations have been mostly focused on single-particle quantum systems, such as, dragged harmonic oscillators \cite{Jarzynski1999b,Talkner2008}, parametric harmonic oscillators \cite{Deffner2008,vanZon2008,Deffner2010,campo_2014}, a single particle in a time-dependent piston \cite{Quan2012}, two-level systems \cite{Quan2008}, and the parametric Morse oscillator \cite{Leonard2014}. However, the interplay of quantum work and quantum statistical properties, e.g., the Fermi-Dirac statistics or the Bose-Einstein statistics have not been fully explored yet (but see Refs.~\cite{Ueda2011,Talkner2011a,Talkner2011b,Talkner2012}). Interference \cite{Tichy2014,Richter2014} of identical particles will undoubtedly influence the thermodynamic properties of many-particle systems.

The difference between the Bose-Einstein distribution and the Fermi-Dirac distribution for identical particles in single-particle eigenstates can be interpreted as a manifestation of the ``static" effect of the interference. Furthermore, in nonequilibrium processes, the transition probabilities between many-particle eigenstates for Bosons and Fermions exhibit interference, as well. This effect can be regarded as the ``dynamic" effect of interference, which profoundly influences the work distribution in nonequilibrium processes.

In this article, we extend our previous studies \cite{Deffner2008,Deffner2010_prl,Deffner2010,Quan2012,Deffner2013} to multi-particle systems. We will show that for noninteracting particles, the transition amplitudes between many-particle eigenstates can be constructed from the transition amplitudes between single-particle eigenstates. From these we obtain the work distribution for arbitrary far-from-equilibrium processes. In practice, however, we will see that for Fermions the work distribution function is relatively easy to compute, whereas for Bosons, the work distribution function is mathematically more involved.

Our findings will be illustrated by two exactly solvable examples -- identical particles confined by a quantum piston and by a harmonic potential. For these model systems we will highlight the significant difference between Bosons and Fermions at low temperatures. On the other hand, in the limit of high temperatures and slow driving we will rediscover the work distribution function for classical particles \cite{Crooks2007}.

Only recently, a ``correspondence principle" for work distributions \cite{Jarzynski2014} has been proposed, which indicates that the quantum distribution converges towards the classical distribution in the semiclassical limit $\hbar\to 0$. Motivated by this result we demonstrate analogously that in the high temperature limit $\beta \to 0$, the work distribution functions for both Bosons and Fermions converge towards that of classical distinguishable particles, which has been previously seen in the single particle case \cite{Deffner2008,Deffner2010,Quan2012,Deffner2011,Deffner2013_epl}. In other words, we demonstrate that in the limit of high temperature, $\beta \to 0$, the quantum work distribution obeys a ``correspondence principle" in the quantum statistical sense independent of the nature of the particles (in contrast to the ``correspondence principle" in the quantum mechanical sense, where $\hbar\to 0$ is required).

Finally, we emphasize that we restrict ourselves to  noninteracting, spinless
identical particles and the nonrelativistic regime. In particular, the particle number is conserved which corresponds to a canonical ensemble. Similar systems have also been studied by Nakamura and his collaborators in Refs. \cite{Nakamura2011,Nakamura2012}, but for grandcanonical ensembles, and their focus has been on averaged quantities rather than on fluctuations.

The paper is organized as follows: In Sec.~\ref{sec:work}, we construct the work distribution for multi-particle systems undergoing nonequilibrium processes. Our findings are illustrated with a 1D box system and a 1D harmonic oscillator, and we numerically compute the work distribution. Section~\ref{sec:conv} is dedicated to the convergence of the quantum work distribution for noninteracting Bosons and Fermions in the limit of high temperature. Finally, Sec.~\ref{sec:con} concludes the discussion with remarks on various properties of the work distribution.

\section{\label{sec:work}Quantum work distribution for identical particles}

In the study of quantum processes operating far from thermal equilibrium one of the key quantities is the work distribution. To the best of our knowledge, previous analyses of multi-particle systems have been restricted to quasistatic processes \cite{Crooks2007}, classical distinguishable particles \cite{Lua2005}, compression of an infinitely large piston \cite{Bena2005}, or sudden quenches of spin chains \cite{Silva2008,Smacchia2013,Fusco2014a}. In particular, analytical results for the work distribution are only known for classical particles \cite{Lua2005,Crooks2007,Bena2005}, whereas the effect of quantum interference is yet to be explored.

In the following we will explicitly construct the quantum work distribution, $\mc{P}(W)$, for systems of many noninteracting particles  (identical or distinguishable), while special focus will be put on the effect of interference on $\mc{P}(W)$. To this end, we have to evaluate the transition probabilities between many-particle eigenstates \cite{Scheel2004,Tichy2012,Tichy2014}, first. In a second step we will illustrate our findings numerically for two simple model systems, namely a 1D piston system and a 1D harmonic oscillator.

\subsection{General expression}

Consider a system of $N$ noninteracting identical particles (either Bosons or Fermions) in a 1D potential. Let us denote the multi-particle eigenstates at the initial and the final instants of a process by $\left| \{i_{k}^{\lambda_{0}}: n_{i_{k}}\}\right\rangle$ and $\left| \{f_{l}^{\lambda_{\tau}}: n_{f_{l}}\}\right\rangle$. Here $\lambda_{0}$ and $\lambda_{\tau}$ denote the initial and the final value of a {\it work parameter} with $\lambda(0)=\lambda_{0}$ and $\lambda(\tau)=\lambda_{\tau}$; $i_{k}^{\lambda_{0}}$ ($f_{l}^{\lambda_\tau}$) is the quantum number of the single-particle state and $n_{i_{k}}$ ($n_{f_{l}}$) is the occupation number of the particles in the $i_{k}$th ($f_{l}$th) eigenstate.

Commonly \cite{Jarzynski1997a,Jarzynski1997b,Crooks1998,Crooks1999},  the system under study is initially prepared in a thermal state at inverse temperature $\beta$, which corresponds here to a canonical ensemble. Then the initial probability to find the system in state $\left| \left\{i_{k}^{\lambda_{0}}:n_{i_{k}} \right\}\right\rangle$ is given by
\begin{equation}
P\left(\left| \left\{i_{k}^{\lambda_{0}}:n_{i_{k}} \right\}\right\rangle\right)=\frac{1}{Z^{\lambda_{0}}}\exp{ \left[ -\beta \left(\sum_{k} n_{i_{k}}E_{i_{k}}^{\lambda_{0}} \right)\right] },
\label{probability}
\end{equation}
where the partition function $Z^{\lambda_{0}}$ reads
\begin{equation}
Z^{\lambda_{0}}=\sum_{\left\{i_{k}:n_{i_{k}}\right\}}\exp{\left[ -\beta \left(\sum_{k} n_{i_{k}}E_{i_{k}}^{\lambda_{0}} \right)\right] }.
\end{equation}
Here we observe the first effect of the quantum statistics. For Fermions we have $n_{i_{k}} \equiv 1$, $\forall k$, due to the Pauli exclusion principle, whereas for Bosons $n_{i_{k}}$ can be an arbitrary positive integer with $n_{i_{k}}\leq N$. The total number of particles, however, is conserved in either case, and we have $\sum_{k} n_{i_{k}}=N$. Finally, $E_{i_{k}}^{\lambda_{0}}$ denotes the $i_{k}$th initial eigenenergy.

After the preparation of the system a projective energy measurement is performed. Then, the external control parameter $\lambda_t$ is varied according to some protocol with $\lambda_{t=0}=\lambda_0$ and $\lambda_{t=\tau}=\lambda_\tau$, and the total system evolves under unitary dynamics. At $t=\tau$ a second projective energy measurement is performed, which induces the system to ``collapse'' into a final multi-particle eigenstate $ \left| \left\{f_{l}^{\lambda_{\tau}}: n_{f_{l}}\right\}\right\rangle$ \cite{Kafri2012}. The work performed during one realization of this protocol is given by
\begin{equation}
\begin{split}
& W\left(\left| \left\{i_{k}^{\lambda_{0}}:n_{i_{k}} \right\}\right\rangle \to \left| \left\{f_{l}^{\lambda_{\tau}}: n_{f_{l}}\right\}\right\rangle \right)=\\
& \quad\quad\sum_{l}n_{f_{l}} E_{f_{l}}^{\lambda_{\tau}} -\sum_{k}n_{i_{k}} E_{i_{k}}^{\lambda_{0}},
\end{split}
\end{equation}
and we denote by $P \left(\left| \left\{i_{k}^{\lambda_{0}}:n_{i_{k}} \right\}\right\rangle \to \left|\left\{f_{l}^{\lambda_{\tau}}: n_{f_{l}}\right\} \right\rangle\right)$  the transition probabilities between many particle eigenstates. Thus, the work distribution,
\begin{equation}
\mc{P}(W)=\la\delta\left[W-W\left(\left| \left\{i_{k}^{\lambda_{0}}:n_{i_{k}} \right\}\right\rangle \to \left| \left\{f_{l}^{\lambda_{\tau}}: n_{f_{l}}\right\}\right\rangle \right)\right]\ra
\end{equation}
can be written as \cite{Tasaki2000,Kurchan2000,Talkner2007}
\begin{widetext}
\begin{equation}
\begin{split}
\mc{P}(W)=\sum_{\left\{i_{k}: n_{i_{k}}\right\}} \sum_{\left\{f_{l}: n_{f_{l}}\right\}} P\left(\left| \left\{i_{k}^{\lambda_{0}}:n_{i_{k}} \right\}\right\rangle\right)
\, P \left(\left|\left\{i_{k}^{\lambda_{0}}:n_{i_{k}} \right\}\right\rangle \to \left |\left\{f_{l}^{\lambda_{\tau}}: n_{f_{l}}\right\}\right\rangle \right) \delta\left[W-\left( \sum_{l}n_{f_{l}} E_{f_{l}}^{\lambda_{\tau}} -\sum_{k}n_{i_{k}} E_{i_{k}}^{\lambda_{0}}\right) \right].
\end{split}
\label{workdistribution}
\end{equation}
\end{widetext}
The latter expression clearly indicates that to calculate the quantum work distribution expressions for the transition probabilities are necessary. Luckily this quantity has been studied extensively in recent years \cite{Scheel2004,Tichy2012,Tichy2014}, and we will here briefly review how to construct the transition probabilities for distinguishable particles, $P_{D}\left( \left| \{i_{k}^{\lambda_{0}}\}\right\rangle \to \left| \{f_{k}^{\lambda_{\tau}}\}\right\rangle \right)$ ($i_{k}^{\lambda_{0}}$ and $f_{k}^{\lambda_{\tau}}$ denote the quantum number of the initial and the final states of the $k$th particle, respectively), as well as for Bosons and for Fermions, $P_{B/F}\left( \left| \{i_{k}^{\lambda_{0}}: n_{i_{k}}\}\right\rangle \to \left| \{f_{l}^{\lambda_{\tau}}: n_{f_{l}}\}\right\rangle \right)$, \cite{Scheel2004,Tichy2012,Tichy2014}.

If the transition amplitude between single particle eigenstates can be expressed as $\left\langle f_{l}^{\lambda_{\tau}} \right | \hat{U} \left | i_{k}^{\lambda_{0}}\right \rangle $, where $\hat{U}$ is the unitary evolution operator corresponding to the Schr\"odinger equation $i \hbar\, \partial_{t} \hat{U}=H(t) \hat{U} $, the transition probabilities between multi-particle eigenstates can be written as \cite{Scheel2004,Tichy2012,Tichy2014}
\begin{widetext}
\begin{equation}
\begin{split}
&P_{D}\left( \left| \{i_{k}^{\lambda_{0}}\}\right\rangle \to \left| \{f_{k}^{\lambda_{\tau}}\}\right\rangle \right)=\prod_{k=1}^{N} \left|\left\langle f_{k}^{\lambda_{\tau}} \right | \hat{U} \left | i_{k}^{\lambda_{0}}\right \rangle \right|^{2},\,\,\, (k=1,2,\cdots,N),\\
&P_{B/F}\left( \left| \{i_{k}^{\lambda_{0}}: n_{i_{k}}\}\right\rangle \to \left| \{f_{l}^{\lambda_{\tau}}: n_{f_{l}}\}\right\rangle \right)= \prod_{l=1}^{L} \frac{1}{n_{f_{l}}!} \prod_{k=1}^{K} \frac{1}{n_{i_{k}}!}\times\\
& \left \vert \left \vert
    \begin{tabular}  {c c c c c c c c}
    $\left\langle f_{1}^{\lambda_{\tau}} \right | \hat{U} \left | i_{1}^{\lambda_{0}}\right \rangle $  & $\cdots$ & $\left\langle f_{1}^{\lambda_{\tau}} \right | \hat{U} \left | i_{1}^{\lambda_{0}}\right \rangle $ & $\left\langle f_{1}^{\lambda_{\tau}} \right | \hat{U} \left | i_{2}^{\lambda_{0}}\right \rangle $ & $\cdots$ & $\left\langle f_{1}^{\lambda_{\tau}} \right | \hat{U} \left | i_{2}^{\lambda_{0}}\right \rangle $& $\cdots$ & $\left\langle f_{1}^{\lambda_{\tau}} \right | \hat{U} \left | i_{K}^{\lambda_{0}}\right \rangle $\\
    $\cdots$ & $\cdots$ & $\cdots$ & $\cdots$ &$\cdots$ & $\cdots$ & $\cdots$ & $\cdots$ \\
    $\left\langle f_{1}^{\lambda_{\tau}} \right | \hat{U} \left | i_{1}^{\lambda_{0}}\right \rangle $  & $\cdots$ & $\left\langle f_{1}^{\lambda_{\tau}} \right | \hat{U} \left | i_{1}^{\lambda_{0}}\right \rangle $ & $\left\langle f_{1}^{\lambda_{\tau}} \right | \hat{U} \left | i_{2}^{\lambda_{0}}\right \rangle $ & $\cdots$ & $\left\langle f_{1}^{\lambda_{\tau}} \right | \hat{U} \left | i_{2}^{\lambda_{0}}\right \rangle $& $\cdots$ & $\left\langle f_{1}^{\lambda_{\tau}} \right | \hat{U} \left | i_{K}^{\lambda_{0}}\right \rangle $\\
    $\left\langle f_{2}^{\lambda_{\tau}} \right | \hat{U} \left | i_{1}^{\lambda_{0}}\right \rangle $ & $\cdots$ & $\left\langle f_{2}^{\lambda_{\tau}} \right | \hat{U} \left | i_{1}^{\lambda_{0}}\right \rangle $ & $\left\langle f_{2}^{\lambda_{\tau}} \right | \hat{U} \left | i_{2}^{\lambda_{0}}\right \rangle $ & $\cdots$ & $\left\langle f_{2}^{\lambda_{\tau}} \right | \hat{U} \left | i_{2}^{\lambda_{0}}\right \rangle $& $\cdots$  & $\left\langle f_{2}^{\lambda_{\tau}} \right | \hat{U} \left | i_{K}^{\lambda_{0}}\right \rangle $\\
      $\cdots$ & $\cdots$ & $\cdots$ & $\cdots$ &$\cdots$ & $\cdots$ & $\cdots$ & $\cdots$ \\
    $\left\langle f_{2}^{\lambda_{\tau}} \right | \hat{U} \left | i_{1}^{\lambda_{0}}\right \rangle $ & $\cdots$ & $\left\langle f_{2}^{\lambda_{\tau}} \right | \hat{U} \left | i_{1}^{\lambda_{0}}\right \rangle $ & $\left\langle f_{2}^{\lambda_{\tau}} \right | \hat{U} \left | i_{2}^{\lambda_{0}}\right \rangle $ & $\cdots$ & $\left\langle f_{2}^{\lambda_{\tau}} \right | \hat{U} \left | i_{2}^{\lambda_{0}}\right \rangle $& $\cdots$  & $\left\langle f_{2}^{\lambda_{\tau}} \right | \hat{U} \left | i_{K}^{\lambda_{0}}\right \rangle $\\
    $\cdots$ & $\cdots$ & $\cdots$ & $\cdots$ &$\cdots$ & $\cdots$ & $\cdots$ & $\cdots$ \\
    $\left\langle f_{L}^{\lambda_{\tau}} \right | \hat{U} \left | i_{1}^{\lambda_{0}}\right \rangle $ & $\cdots$ & $\left\langle f_{L}^{\lambda_{\tau}} \right | \hat{U} \left | i_{1}^{\lambda_{0}}\right \rangle $ & $\left\langle f_{L}^{\lambda_{\tau}} \right | \hat{U} \left | i_{2}^{\lambda_{0}}\right \rangle $  & $\cdots$ & $\left\langle f_{L}^{\lambda_{\tau}} \right | \hat{U} \left | i_{2}^{\lambda_{0}}\right \rangle $ & $\cdots$ & $\left\langle f_{L}^{\lambda_{\tau}} \right | \hat{U} \left | i_{K}^{\lambda_{0}}\right \rangle $\\
    \end{tabular}
  \right\vert_{\zeta} \right\vert^{2},
\end{split}
\label{transitonprobability}
\end{equation}
\end{widetext}
where the matrix element $\left\langle f_{l}^{\lambda_{\tau}} \right | \hat{U} \left | i_{k}^{\lambda_{0}}\right \rangle $ occupies a block of size $n_{f_{l}} \times n_{i_{k}}$. Due to the conservation of the particle number we have as before,
$N=\sum_{l=1}^{L} n_{f_{l}} =\sum_{k=1}^{K} n_{i_{k}}$. Furthermore, $\zeta=-1$ and $\zeta=1$ in Eq.~(\ref{transitonprobability}) correspond to Fermions and Bosons, respectively. For Fermions the transition amplitude is equal to the determinant of the matrix (\ref{transitonprobability}), whereas for Bosons, the transition amplitude is given by the permanent \cite{Scheel2004,Tichy2012,Tichy2014}.

Generally, to compute the transition probabilities merely the transition amplitudes between single particle eigenstates $\left\langle f_{l}^{\lambda_{\tau}} \right | \hat{U} \left | i_{k}^{\lambda_{0}}\right \rangle $ are necessary.
However, we will see shortly that, in practice, the calculation of the permanent, the case of Bosons, is much more involved than the calculation of the determinant, for Fermions.

We now can proceed to compute the  quantum work distribution for specific many-particle systems. For the sake of simplicity we will continue our discussion for two analytically solvable examples. For single-particle systems analogous studies include the 1D piston system with a moving wall \cite{Doescher1969,Quan2012} and the 1D harmonic oscillator with a time-dependent angular frequency \cite{Husimi1953,Deffner2008,Deffner2010}.

\subsection{Case one: particles in one dimensional piston}

\begin{figure}[]%
      \subfigure{%
         \label{fig1}%
         \includegraphics[width=.42\textwidth]{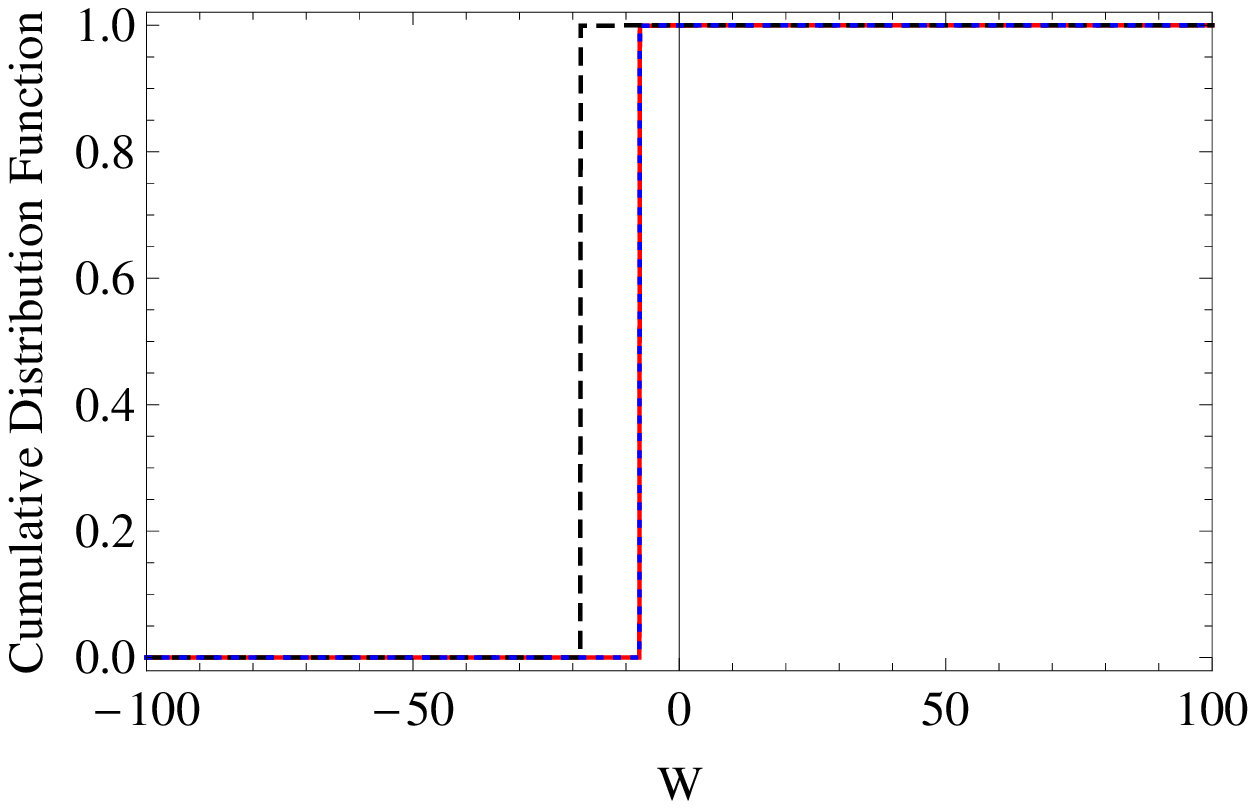}
        }
      \subfigure{%
         \label{fig2}%
        \includegraphics[width=.42\textwidth]{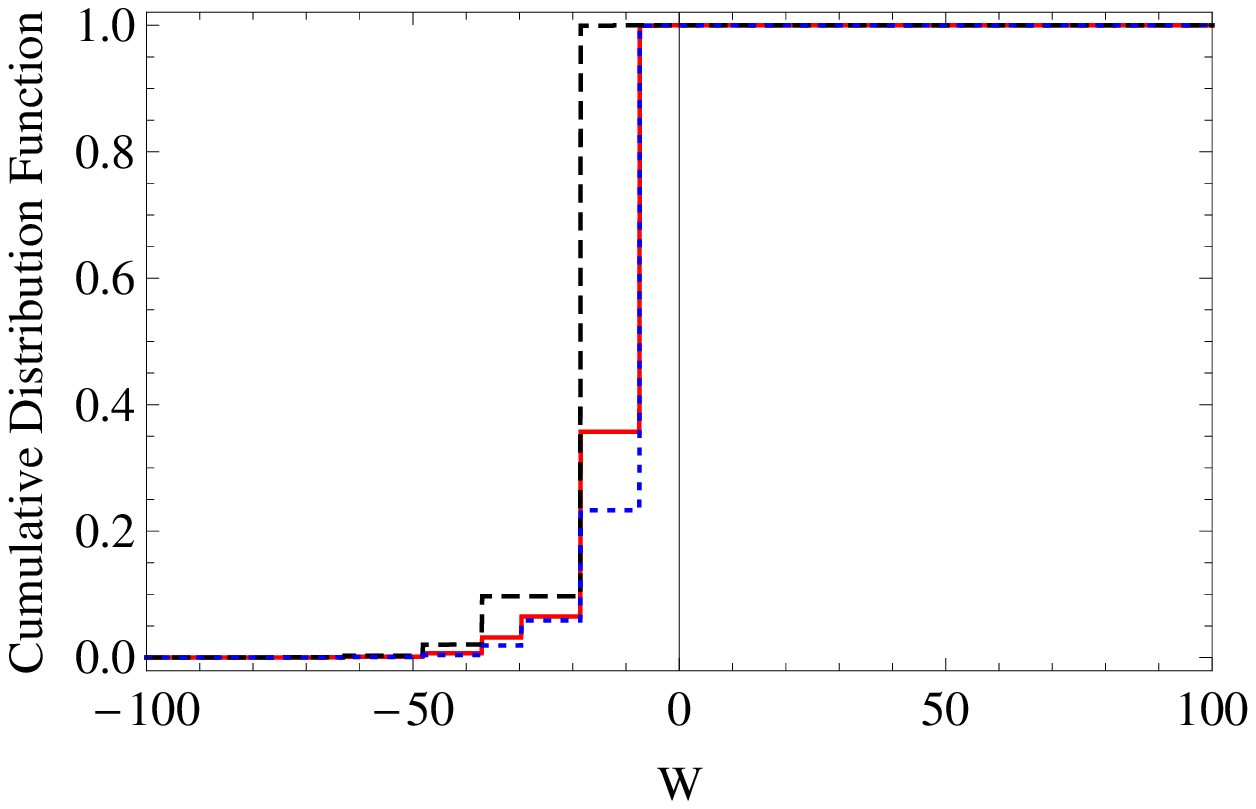}
      }
      \subfigure{%
         \label{fig3}%
        \includegraphics[width=.42\textwidth]{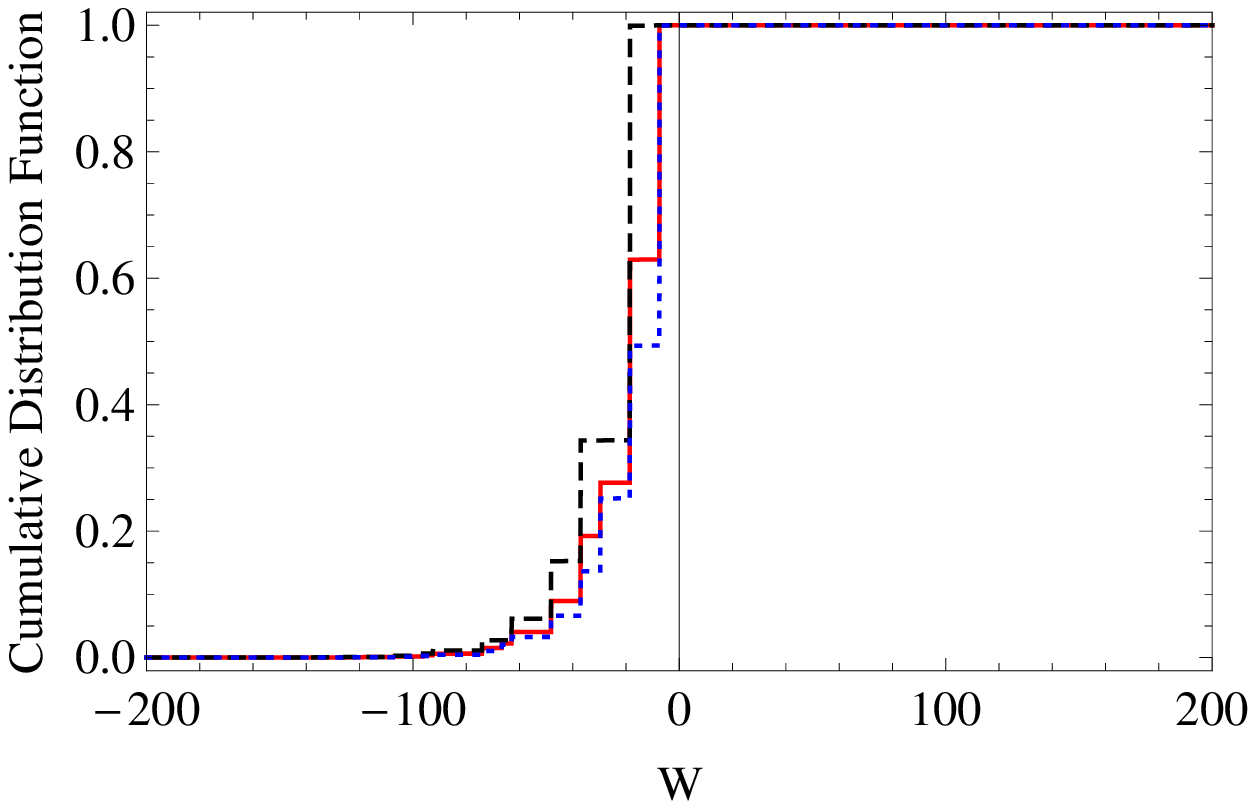}
     }
      \subfigure{%
         \label{fig5}%
        \includegraphics[width=.42\textwidth]{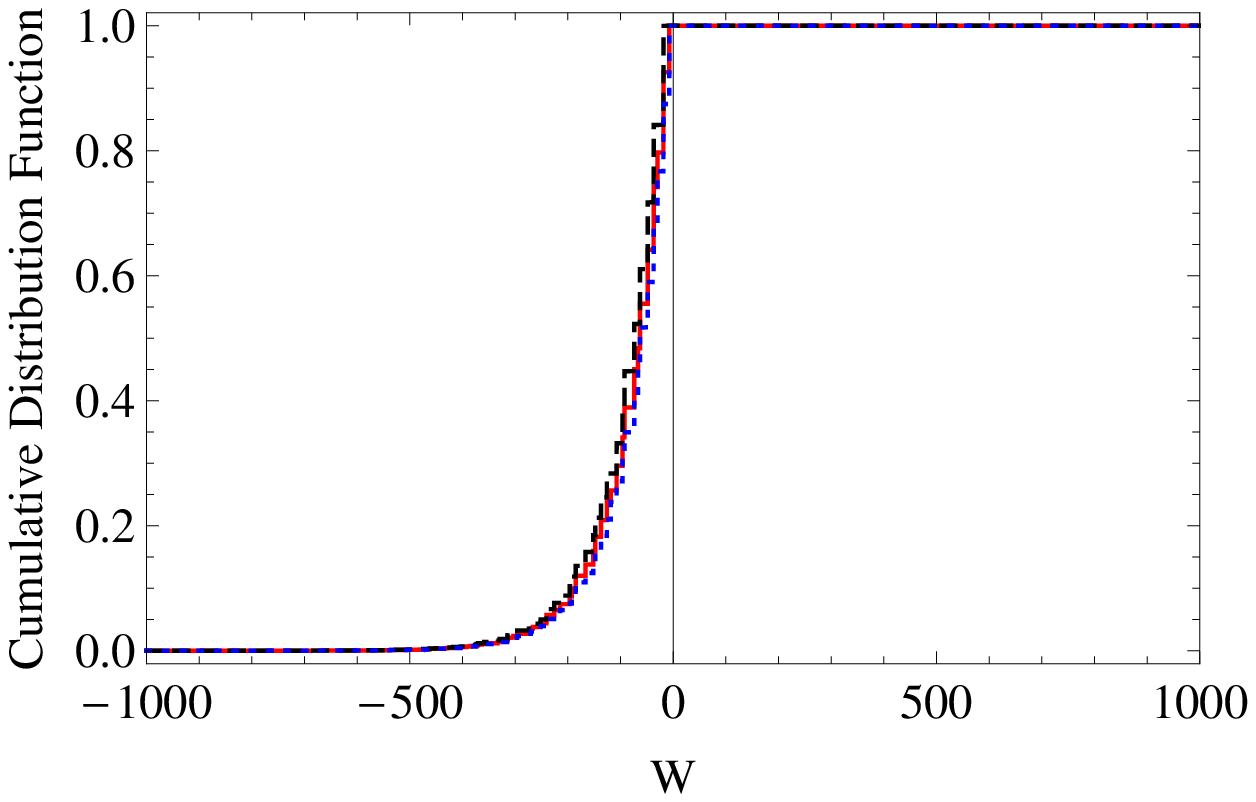}
      }
   \caption{\label{fig:piston_2_slow}(color online) Cumulative work distribution \eqref{eq:cum} for two  Bosons (blue, dotted line), Fermions (black, dashed line) and distinguishable particles (red, solid line) in an expanding piston with $\lambda_{0}=1$, $\lambda_{\tau}=2$ and $v=0.1$. Temperatures are from top to bottom $\beta^{-1}=0$, $\beta^{-1}=10$, $\beta^{-1}=20$, and $\beta^{-1}=100$.}
\end{figure}

\begin{figure}[]%
      \subfigure{%
         \label{fig7}%
        \includegraphics[width=.42\textwidth]{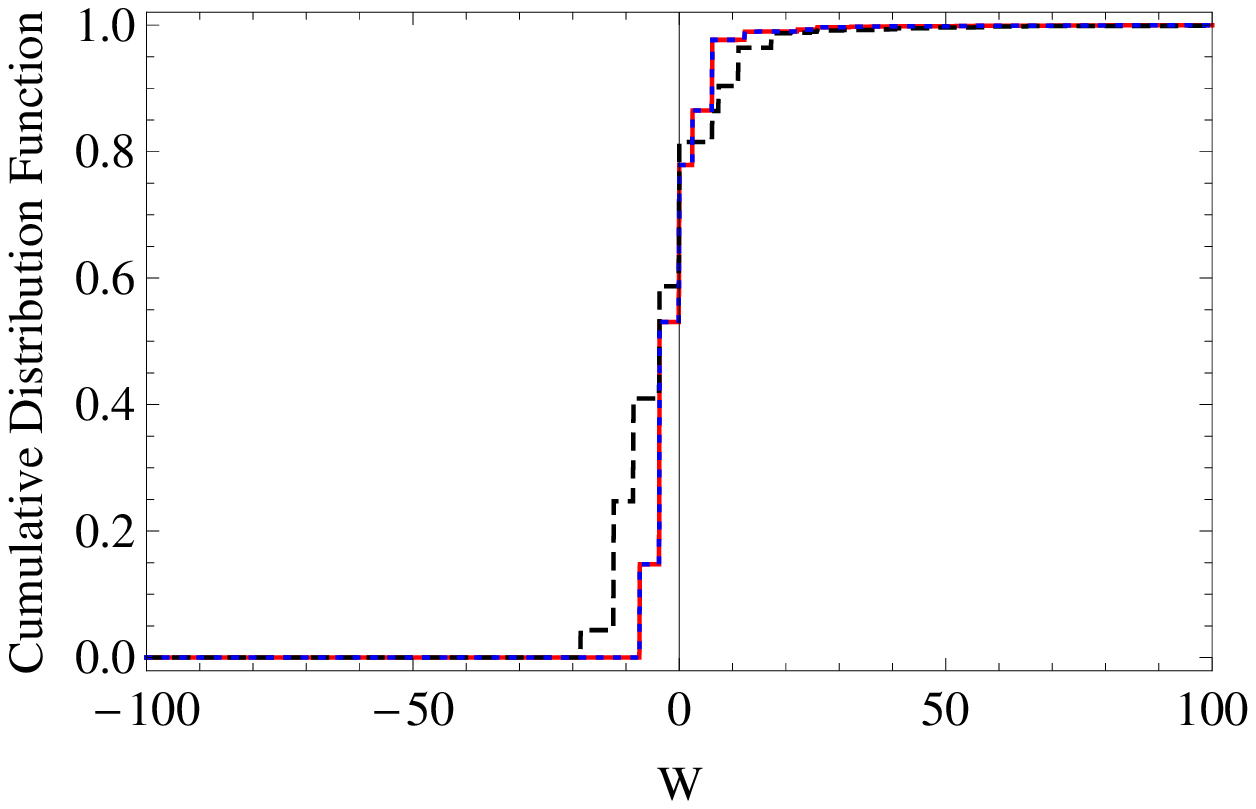}
      }
      \subfigure{%
         \label{fig8}%
        \includegraphics[width=.42\textwidth]{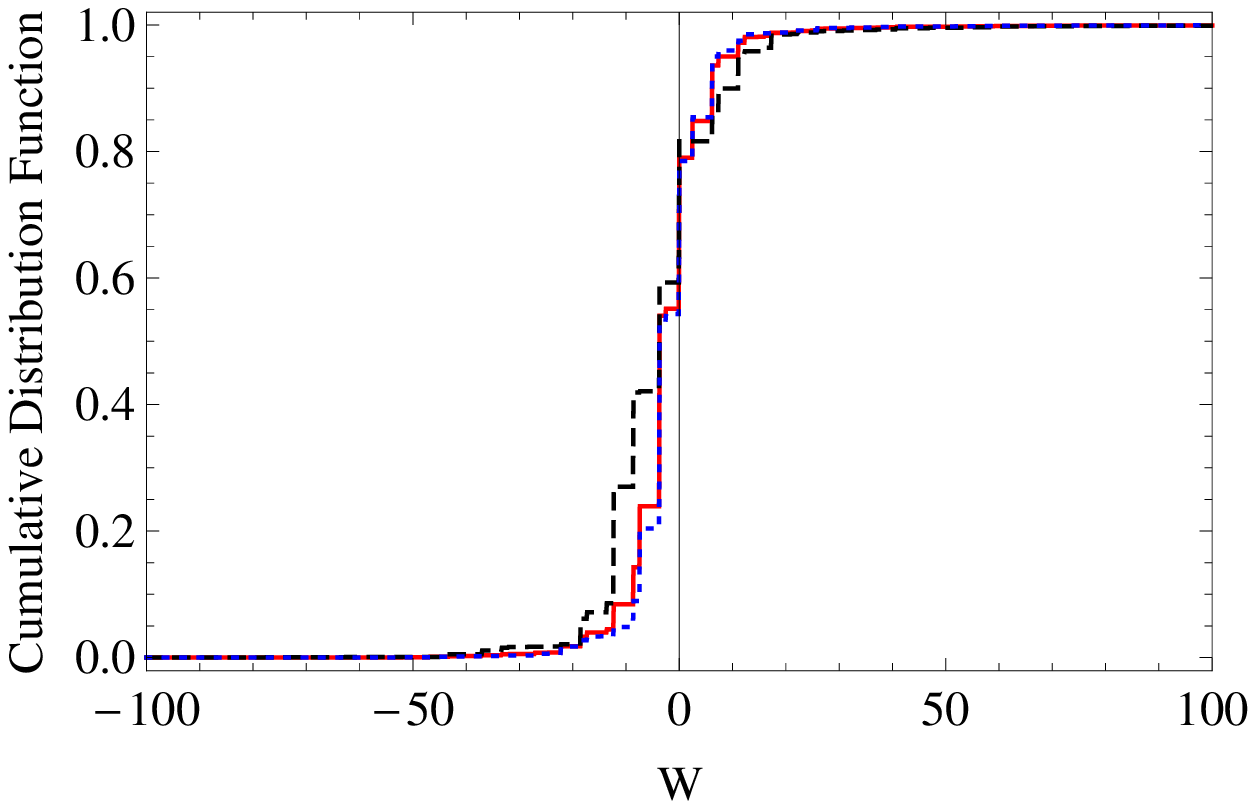}
      }
      \subfigure{%
         \label{fig9}%
        \includegraphics[width=.42\textwidth]{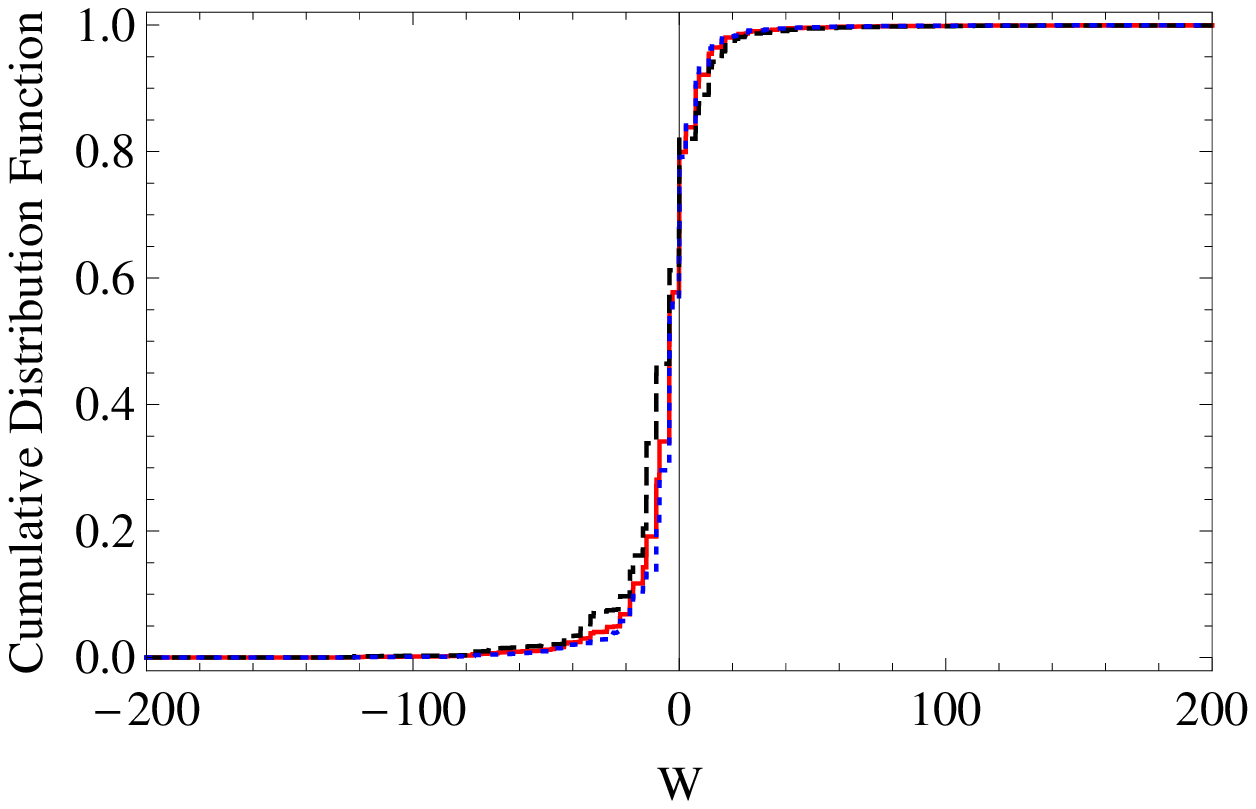}
      }
      \subfigure{%
         \label{fig11}%
        \includegraphics[width=.42\textwidth]{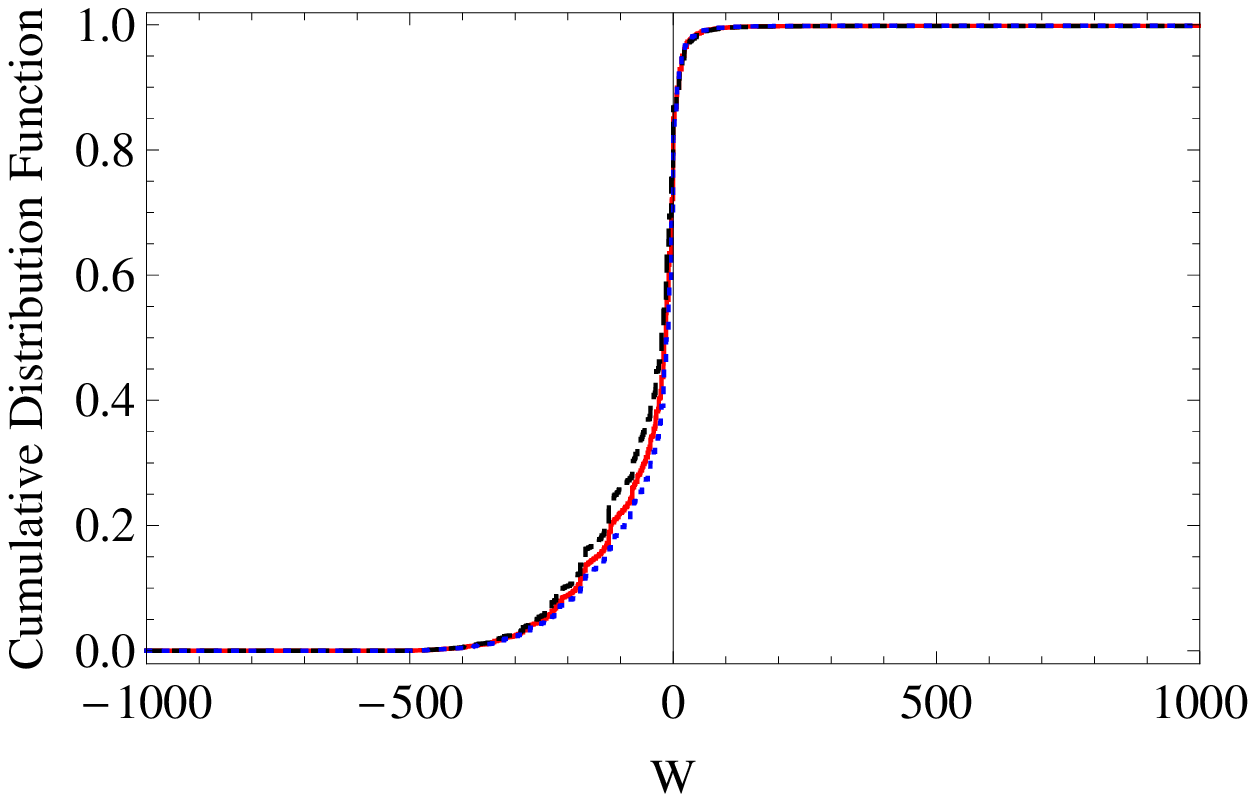}
      }
   \caption{\label{fig:piston_2_intermediate}(color online) Cumulative work distribution \eqref{eq:cum} for two  Bosons (blue, dotted line), Fermions (black, dashed line) and distinguishable particles (red, solid line) in an expanding piston with $\lambda_{0}=1$, $\lambda_{\tau}=2$, and $v=8$. Temperatures are from top to bottom $\beta^{-1}=0$, $\beta^{-1}=10$, $\beta^{-1}=20$, and $\beta^{-1}=100$.}
\end{figure}

The paradigm system in statistical mechanics is undoubtedly the classical ideal gas confined by a piston. Quantum particles in an infinite square well can be considered a quantum analog. The dynamics of a single particle in this ``quantum piston'' has been studied extensively in various contexts, see for instance Refs.~\cite{Doescher1969,Quan2012,Campo2012} and references therein. When the piston is pulled or compressed at a constant velocity, analytical solutions to the transition amplitudes between the initial and the final energy eigenstates $\left\langle f_{l}^{\lambda_{\tau}} \right | \hat{U} \left | i_{k}^{\lambda_{0}}\right \rangle$ can be obtained analytically \cite{Doescher1969,Quan2012}. Specifically, for a quantum piston expanding at a constant velocity $v$ from an initial length $\lambda(0)=\lambda_{0}$, $\lambda(t)=\lambda_{0}+vt$, a set of independent solutions to the time-dependent Schr\"odinger equation can be written as \cite{Doescher1969}
\begin{equation}
\Phi_{j}(x,t)=\exp\left[\frac{i}{\hbar \lambda(t)}\left( \frac{1}{2}Mvx^{2}-E_{j}^{\lambda_{0}}\lambda_{0} t\right) \right] \phi_{j}(x,\lambda(t)),
\end{equation}
where $j=1,2,3,\cdots$ and  $M$ is the mass of the particle, $E_{j}^{\lambda_{0}}=j^{2}\pi^{2}\hbar^{2}/2M\lambda_{0}^{2}$ is the $j$th eigenenergy and $\phi_{j}(x,\lambda)$ is the $j$th energy eigenstate of a particle in an infinite square potential
\begin{equation}
\phi_{j}(x,\lambda) =\sqrt{\frac{2}{\lambda}} \sin { \left( \frac{j \pi x}{\lambda} \right)}.
\end{equation}
A general solution of the time-dependent Schr\"odinger equation takes the form
\begin{equation}
\Psi(x,t)=\sum_{j}^{\infty} c_{j}\Phi_{j}(x,t),
\end{equation}
where the time-independent coefficients $c_{j}$ are set by the initial wave function
\begin{equation}
c_{j}=\int_{0}^{\lambda_{0}}\td x\, \Phi_{j}^{\ast}(x,0)\Psi(x,0).
\end{equation}
For initial conditions $\Psi(x,0)=\phi_{i_{k}}(x,\lambda_{0})\equiv \left\langle x  \vert i_{k}^{\lambda_{0}} \right\rangle$ these coefficients are (setting $\hbar=1$ and $M=1$)
\begin{equation}
\begin{split}
c_{j}(i_{k})&=\frac{2}{\lambda_{0}}\int_{0}^{\lambda_{0}}\td x\, \e{-i\frac{vx^{2}}{2\lambda_{0}}}\\
&\quad\times\sin{\left( \frac{j\pi x}{\lambda_{0}}\right)}\sin{\left( \frac{i_{k}\pi x}{\lambda_{0}}\right)},
\end{split}
\end{equation}
and the time evolution matrix elements to the state $\left | f_{l}^{\lambda_{\tau}}\right\rangle$ at the final instant $t=\tau$ become
\begin{equation}
\left\langle f_{l}^{\lambda_{\tau}} \right | \hat{U} \left | i_{k}^{\lambda_{0}}\right\rangle =\sum_{j=1}^{\infty} c_{j}(i_{k}) \int_{0}^{\lambda_{\tau}}\td x\, \Phi_{j}(x,\tau) \phi^{\ast}_{f_{l}}(x,\lambda_{\tau}).
\label{amplitude}
\end{equation}
Substituting the transition amplitude (\ref{amplitude}) into Eq.~(\ref{transitonprobability}) we obtain the transition probabilities between the multi-particle eigenstates in the piston.

\begin{figure}[]%
      \subfigure{%
         \label{fig13}%
        \includegraphics[width=.42\textwidth]{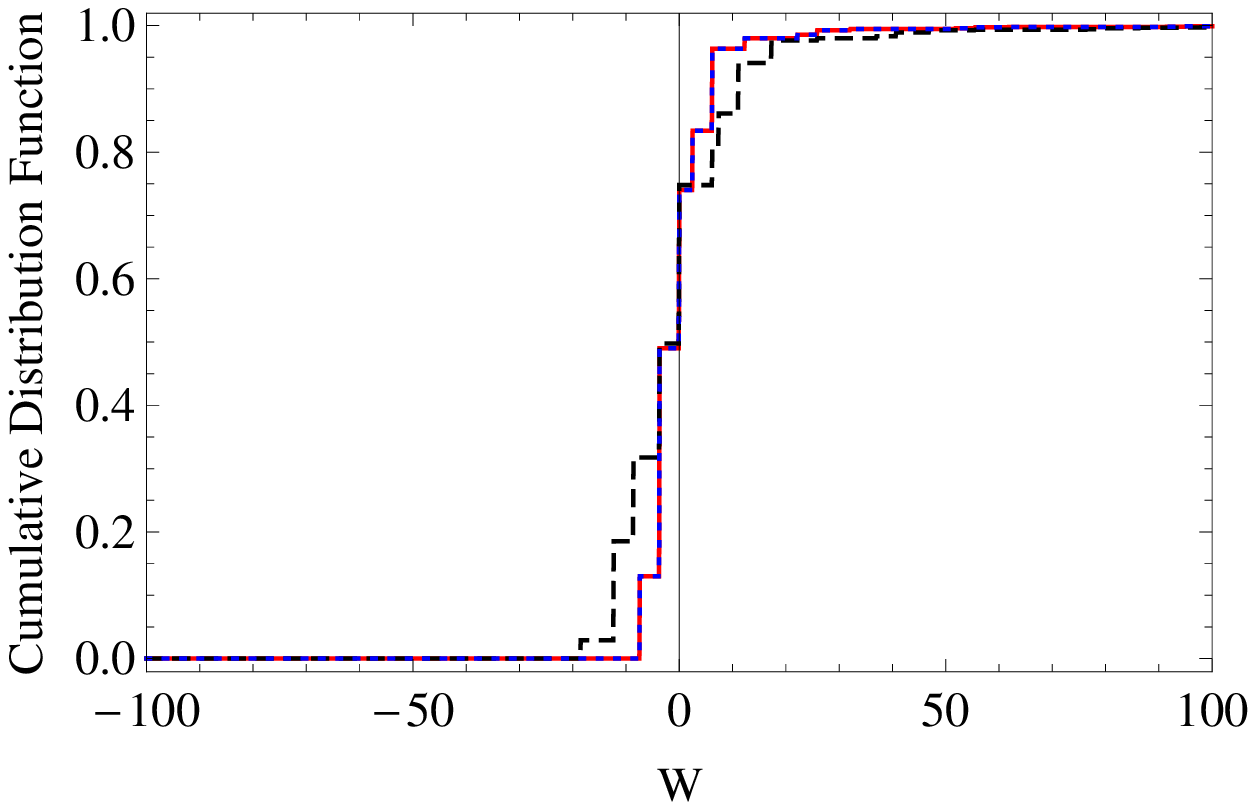}
      }
      \subfigure{%
         \label{fig14}%
        \includegraphics[width=.42\textwidth]{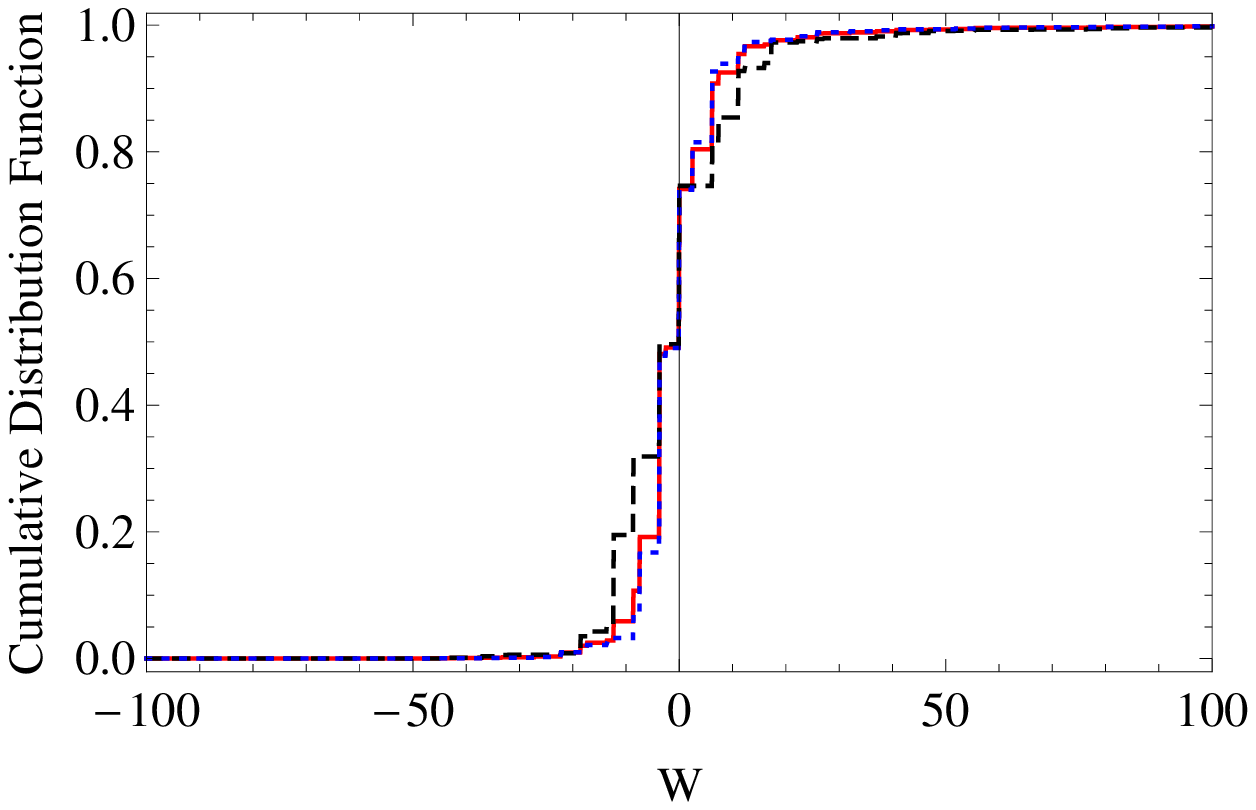}
      }
      \subfigure{%
         \label{fig15}%
        \includegraphics[width=.42\textwidth]{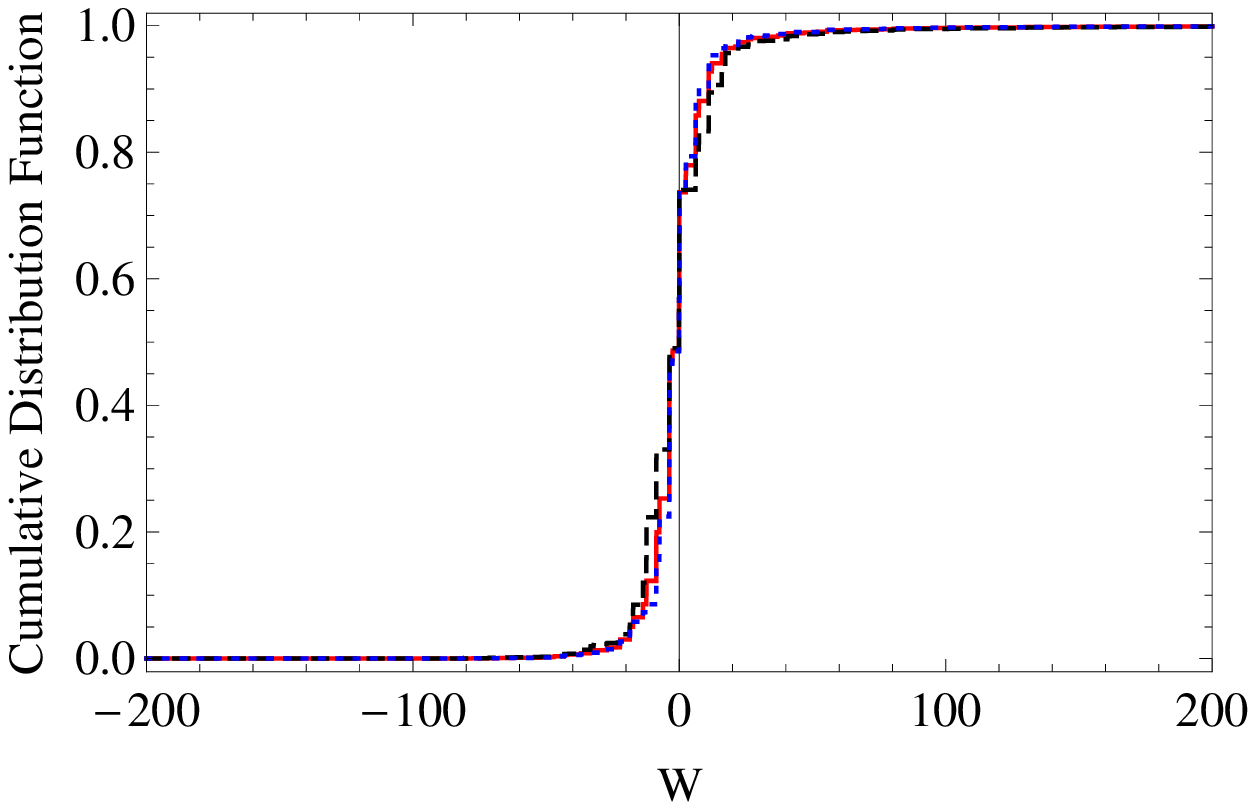}
     }
      \subfigure{%
         \label{fig17}%
        \includegraphics[width=.42\textwidth]{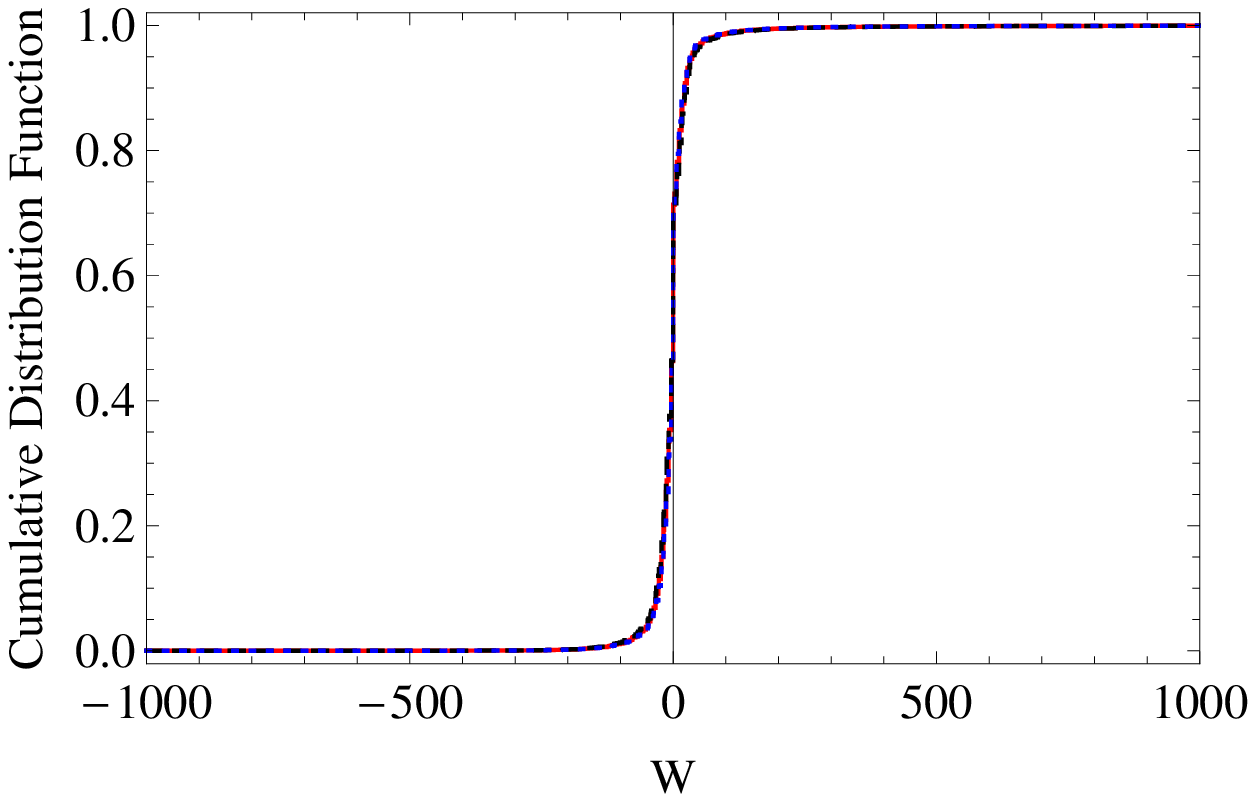}
      }
   \caption{\label{fig:piston_2_fast}(color online) Cumulative work distribution\eqref{eq:cum} for  two  Bosons (blue, dotted line), Fermions (black, dashed line) and distinguishable particles (red, solid line) in an expanding piston with $\lambda_{0}=1$, $\lambda_{\tau}=2$, and $v=100$. Temperatures are from top to bottom $\beta^{-1}=0$, $\beta^{-1}=10$, $\beta^{-1}=20$, and $\beta^{-1}=100$.}
\end{figure}

\begin{figure}[]%
        \subfigure{%
         \label{fig22}%
        \includegraphics[width=.42\textwidth]{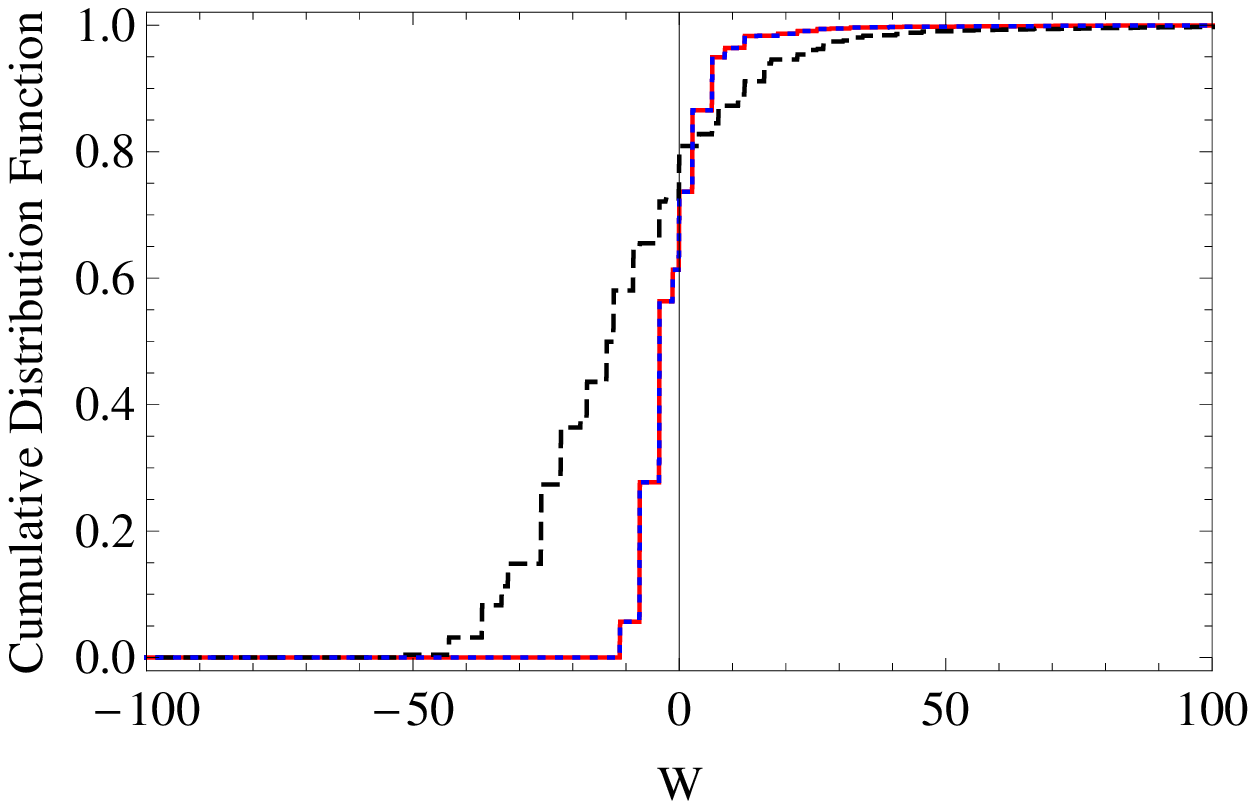}
      }
      \subfigure{%
         \label{fig23}%
        \includegraphics[width=.42\textwidth]{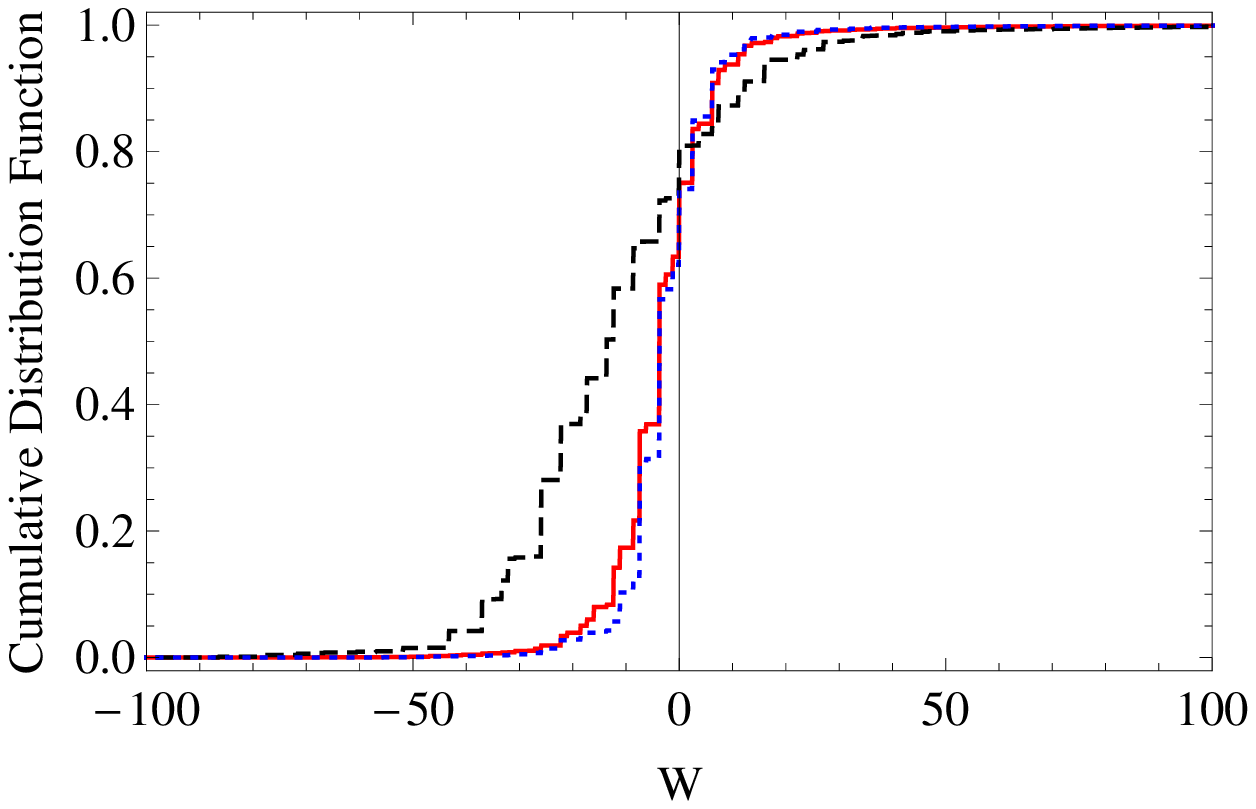}
      }
      \subfigure{%
         \label{fig24}%
        \includegraphics[width=.42\textwidth]{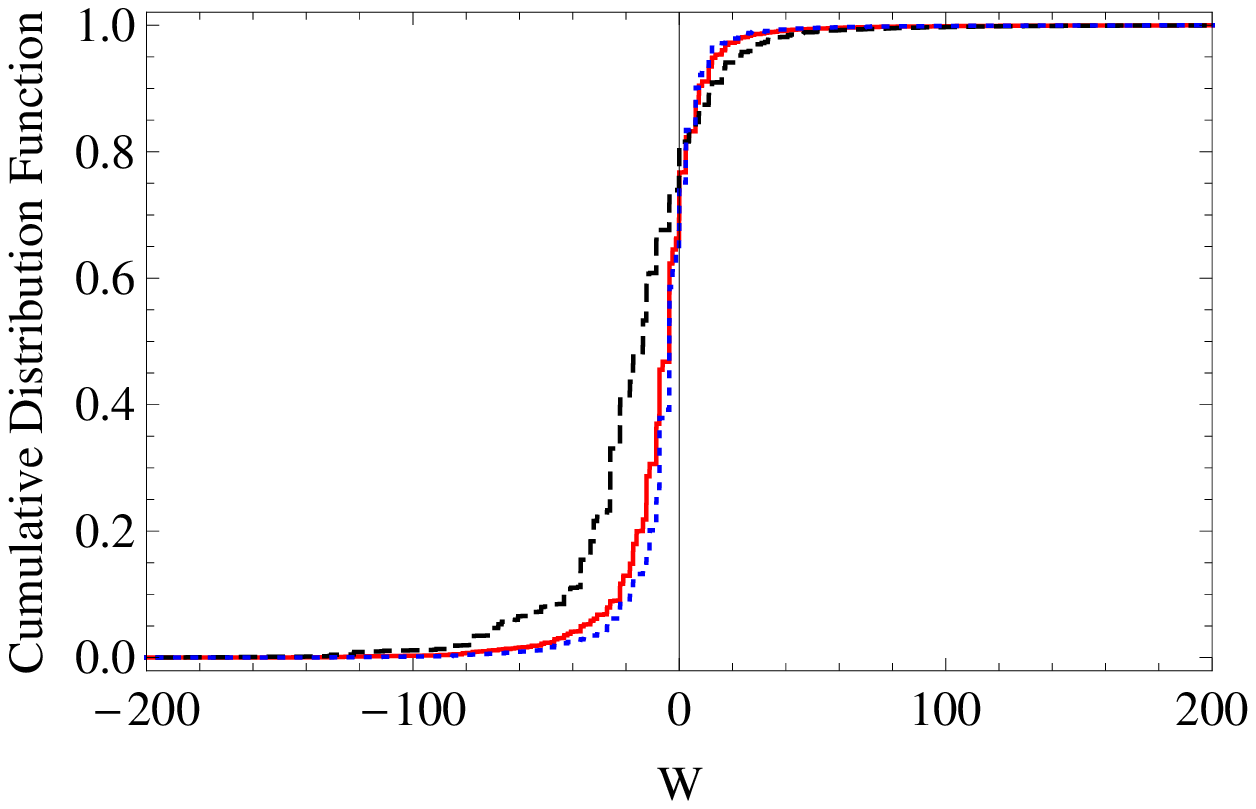}
      }
   \caption{\label{fig:piston_3_slow}(color online) Cumulative work distribution \eqref{eq:cum} for  three Bosons (blue, dotted line), Fermions (black, dashed line) and distinguishable particles (red, solid line) in an expanding piston with $\lambda_{0}=1$, $\lambda_{\tau}=2$, and $v=8$. Temperatures are from top to bottom $\beta^{-1}=0$, $\beta^{-1}=10$, and $\beta^{-1}=20$. }
\end{figure}

\begin{figure}[]%
\begin{center}
      \subfigure{%
         \label{fig25}%
        \includegraphics[width=.42\textwidth]{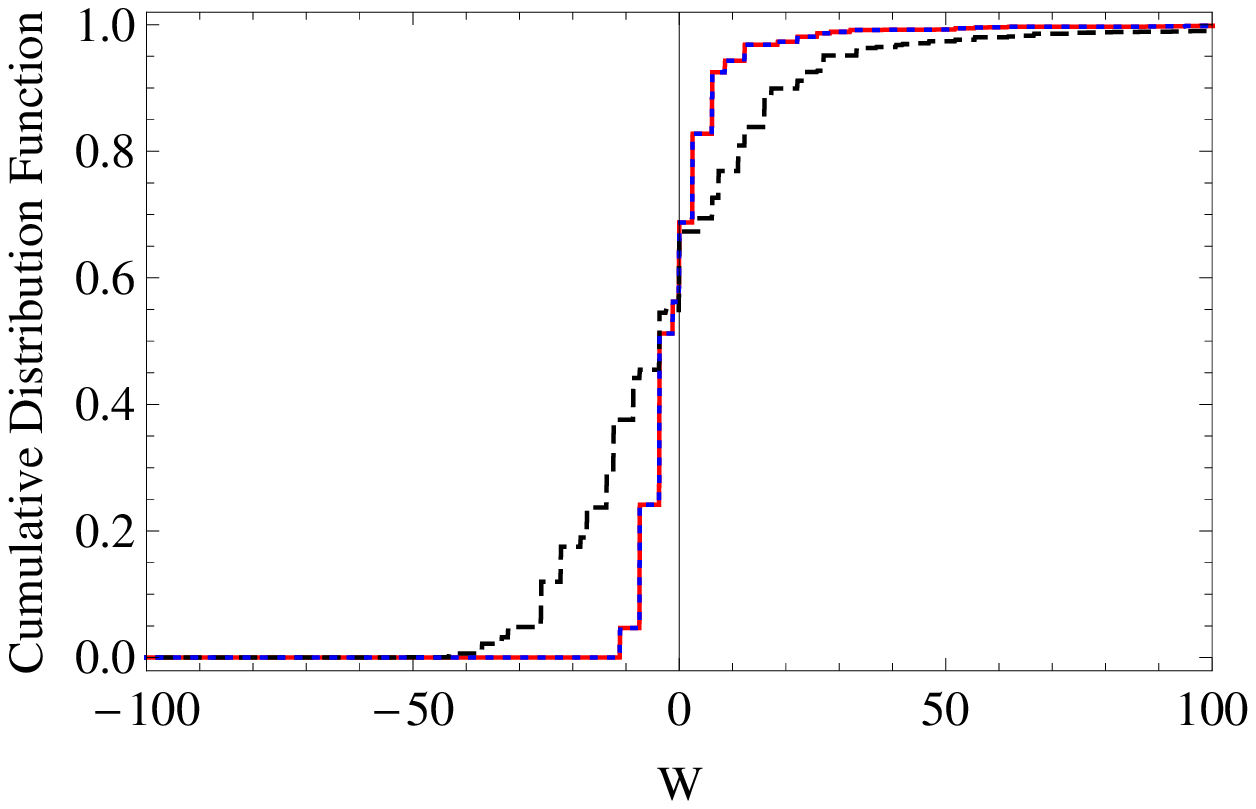}
      }
      \subfigure{%
         \label{fig26}%
        \includegraphics[width=.42\textwidth]{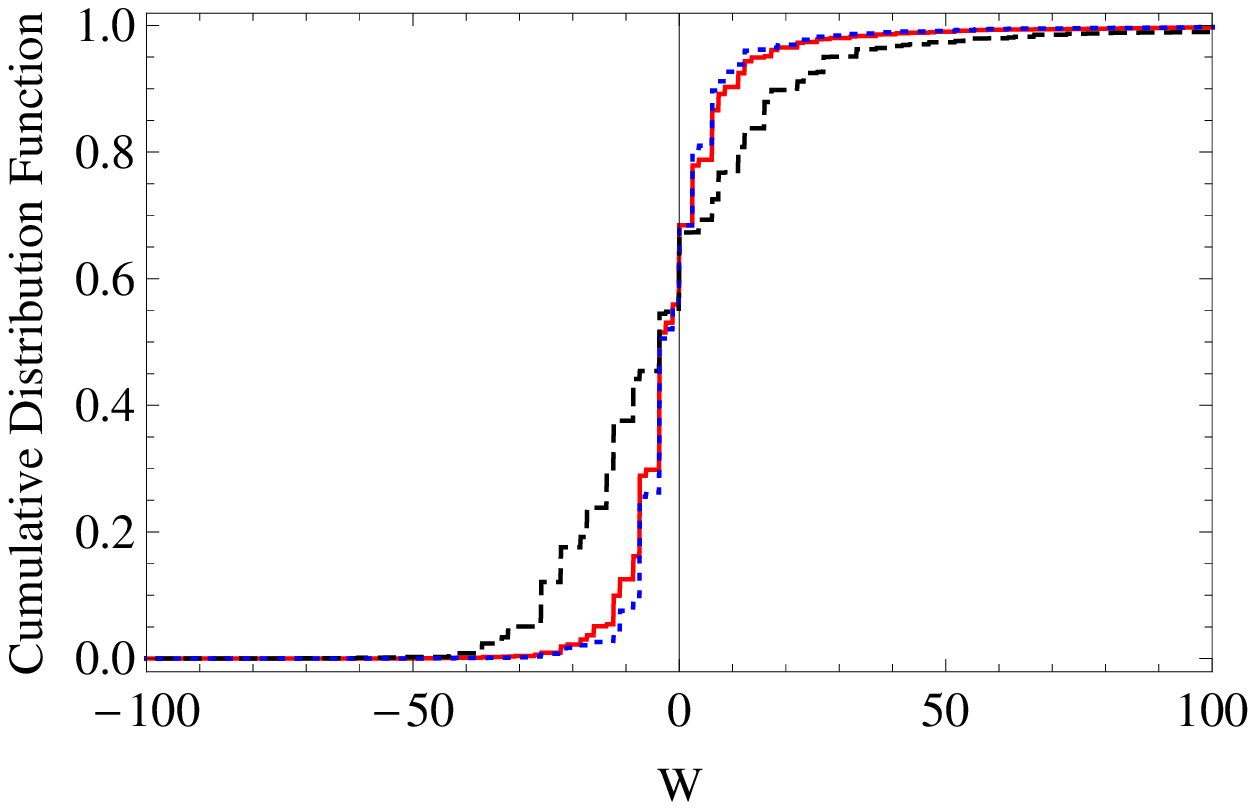}
      }
      \end{center}
   \caption{\label{fig:piston_3_fast}(color online) Cumulative work distribution \eqref{eq:cum} for  three Bosons (blue, dotted line), Fermions (black, dashed line) and distinguishable particles (red, solid line) in an expanding piston with $\lambda_{0}=1$, $\lambda_{\tau}=2$ and $v=100$. Temperatures are from top to bottom $\beta^{-1}=0$ and $\beta^{-1}=10$.}
\end{figure}

We plot numerical results of the work distribution for the case of two or three identical particles in Figs.~\ref{fig:piston_2_slow}-\ref{fig:piston_3_fast}. For the sake of clarity we plot the cumulative distributions,
\begin{equation}
\label{eq:cum}
\rho(W)=\int^W\td W'\,\mc{P}(W')
\end{equation}
rather than the quantum work distributions $\mc{P}(W)$. In all figures we compare the results for distinguishable (Boltzmann) particles (red lines) with those for Fermions (black lines) and for Bosons (blue lines).

We start with the case of slow expansion in Fig.~\ref{fig:piston_2_slow} with $v=0.1$. We observe that in the limit of low temperature the work distributions for Bosons and for distinguishable particles are identical. This can be understood by noting that (i) at $T=0$ both Bosons and distinguishable particles occupy only the single-particle ground state, and (ii) at $T=0$ the transition probabilities between many-particle states for Bosons and for distinguishable particles are identical (all the transition probabilities for distinguishable particles, which correspond to the same state for Bosons, should be summed up). We also observe that in the limit of low temperature, the work distributions for Bosons and for Fermions differ significantly due to the static interference of identical particles.

At intermediate temperatures, e.g., from $\beta^{-1}=10$ to $\beta^{-1}=20$, the work distribution function for distinguishable particles locates between that for Bosons and that for Fermions. In contrast to distinguishable particles, there is an effective ``attractive" interaction among Bosons, while there is an effective ``repelling" interaction among Fermions. As a result, Fermions perform more work than distinguishable particles on the piston during an expanding process, while Bosons perform less. By further increasing the initial temperature, the cumulative work distribution functions become smoother and smoother, and the work distribution functions for Bosons and for Fermions show a tendency of convergence.

In the limit of high temperature (e.g., in Fig.~\ref{fig:piston_2_slow} $\beta^{-1}=100$ can already be regarded as the limit of high temperature), the work distribution functions for the three kinds of particles collapse onto the same curve. In Figs.~\ref{fig:piston_2_intermediate} and \ref{fig:piston_2_fast}  all parameters are the same as those in Fig.~\ref{fig:piston_2_slow} except that the speed of the expansion of the piston is higher. We observe that the faster the speed of the expansion the faster, i.e., at lower temperature, the convergence of the work distribution functions for the three kinds of particles.  Notice that the convergence depends on both the work protocol and the initial temperature, and that the convergence is generally not uniform. In Fig.~\ref{fig:piston_2_fast} (fast protocol) for the ``typical" values of work, the convergence is faster than that for the work values in the tails of the distribution. However, in Fig.~\ref{fig:piston_2_intermediate} (intermediate protocol) this is not the case. In Fig.~\ref{fig:piston_2_intermediate} we can see that for the ``typical'' values of work, the convergence is slower. In Fig.~\ref{fig:piston_2_slow} (slow protocol) the convergence is approximately uniform.

In Figs.~\ref{fig:piston_3_slow}-\ref{fig:piston_3_fast} we plot the cumulative work distribution \eqref{eq:cum} for the case of three identical particles at low temperatures. The difference of the work distribution functions between Bosons and Fermions for three particles are more prominent than those for two particles. It can be inferred that, with the increase of the particle number, the distinguishability of the work distributions of Bosons and Fermions at low temperature will becomes even more significant. This can be understood by considering that, at low temperature, the system will stay in a state close to the many-particle ground state. For Bosons and Fermions the ground states are a Bose condensate and a Fermi sea, respectively.

By further increasing the particle number, the complexity of the calculation of the transition probabilities between many-particle eigenstates increases exponentially with the particle number. Also, with increase of the temperature, the number of eigenstates, which will be visited during the work process, increases dramatically. Therefore, we restricted ourselves to two and three particles and to rather low temperatures. However, from the work distribution functions for three particles at $\beta^{-1}=0$, $\beta^{-1}=10$ and $\beta^{-1}=20$ (see Figs.~\ref{fig:piston_3_slow}-\ref{fig:piston_3_fast}), one already observes the tendency to converge by raising the temperature.


\subsection{Case two: particles in one dimensional harmonic potential}

As a second case study of pedagogical value we analyze the 1D harmonic oscillator. Specifically, we consider the Hamiltonian
\begin{equation}
\label{eq:harm}
H_{s}(x,t)=\frac{1}{2M}\frac{d^{2}}{dx^{2}}+\frac{1}{2}M \omega_t^2\,x^{2},
\end{equation}
where $M=1$ is again the single particle mass and we identify the work parameter with the angular frequency, $\lambda_t=\omega_t$. This system can be solved analytically, see for instance \cite{Husimi1953,Cervero1999}, and has been developed as the prototypical example in quantum thermodynamics \cite{He2002,Lin2003,Quan2007,Quan2009,Deffner2008,Talkner2008,Huber2008,Wang2009,Wang2011,Wang2012,Deffner2010,Abah2012,Deffner2013,Abah2014_epl,Rossnagel2014,campo_2014}.

It has been shown that the single-particle time evolution operator can be written in position space as,
\begin{equation}
\label{eq:prop}
\begin{split}
U_t(x;\, x_0)&=\sqrt{\frac{M}{2\pi i \hbar\, X_t}}\\
&\quad\times\e{\frac{i M}{2 \hbar X_t}\,\left(\dot{X}_t \,x^2-2 x x_0+Y_t x_0^2\right)}
\end{split}
\end{equation}
where $X_t$ and $Y_t$ are solutions of the classical, force free equation of motion,
\begin{equation}
\label{eq:force}
\ddot{\xi_t}+\omega_t^2\, \xi_t=0
\end{equation}
with $X_{t=0}=0$, $\dot{X}_{t=0}=1$ and $Y_{t=0}=1$, $\dot{Y}_{t=0}=0$. From the latter the single-particle propagator can be obtained in energy representation by evaluating
\begin{equation}
\label{eq:trafo}
\bra{f^{\lambda_\tau}}\hat{U}\ket{i^{\lambda_0}}=\int\td x\int\td x_0\, \psi_{f}(x) \,U_t(x;\, x_0) \psi_i(x_0)\,,
\end{equation}
where $\psi_\nu (x)$, ($\nu=i,f$) are the instantaneous eigenstates of the time-dependent Hamiltonian $H_s(x,t)$ \eqref{eq:harm}. The result is a rather lengthy expression \cite{Deffner2010}, which we summarize in Appendix~\ref{sec:appD}.

In Figs.~\ref{fig:harm_2_slow}-\ref{fig:harm_3_fast} we plot the cumulative work distribution \eqref{eq:cum} for the linear quench,
\begin{equation}
\label{eq:quench}
\omega_t^2=\omega_0^2+\left(\omega_\tau^2-\omega_0^2\right)\,t/\tau\,.
\end{equation}
Generally, the same features as those in the case of the 1D piston system can be observed: In the limit of low temperature, the work distribution functions for Bosons and for Fermions differ significantly, while for high temperatures the three distributions converge. Also, one may notice that in the 1D piston system the cumulative work has contributions in both negative and positive values (see e.g. Figs.~\ref{fig:piston_2_intermediate}-\ref{fig:piston_2_fast}). The negative  (positive) value of work corresponds to the ``trajectory" of jumping from a higher (lower) energy state to a lower (higher) energy state. In the harmonic oscillator, we also see that the cumulative work has a tiny tail for the negative values (see third line of Fig.~\ref{fig:harm_2_fast}), but it is much less prominent than those in the piston system. This is due to the present choice of the quench protocol (see also Fig. 3 of Ref.~\cite{Deffner2010}). One can expect that if we properly choose the initial temperature and the quench speed, that the tail for negative value will become more prominent (see Figs. 1-2 of Ref.~\cite{Deffner2010}).

An interesting feature to note is that the interference seems to play much less of a role for the harmonic oscillator than for the piston. In particular, the work distributions for Bosons, Fermion, and distinguishable particles start converging at much lower temperatures. This can be understood by noticing that the energy levels of the harmonic oscillator are much denser than the ones of the square well potential. Thus, interference effects are ``smeared out'' already at finite but low temperatures.

\begin{figure}[]%
      \subfigure{%
         \label{fig5a}%
        \includegraphics[width=.42\textwidth]{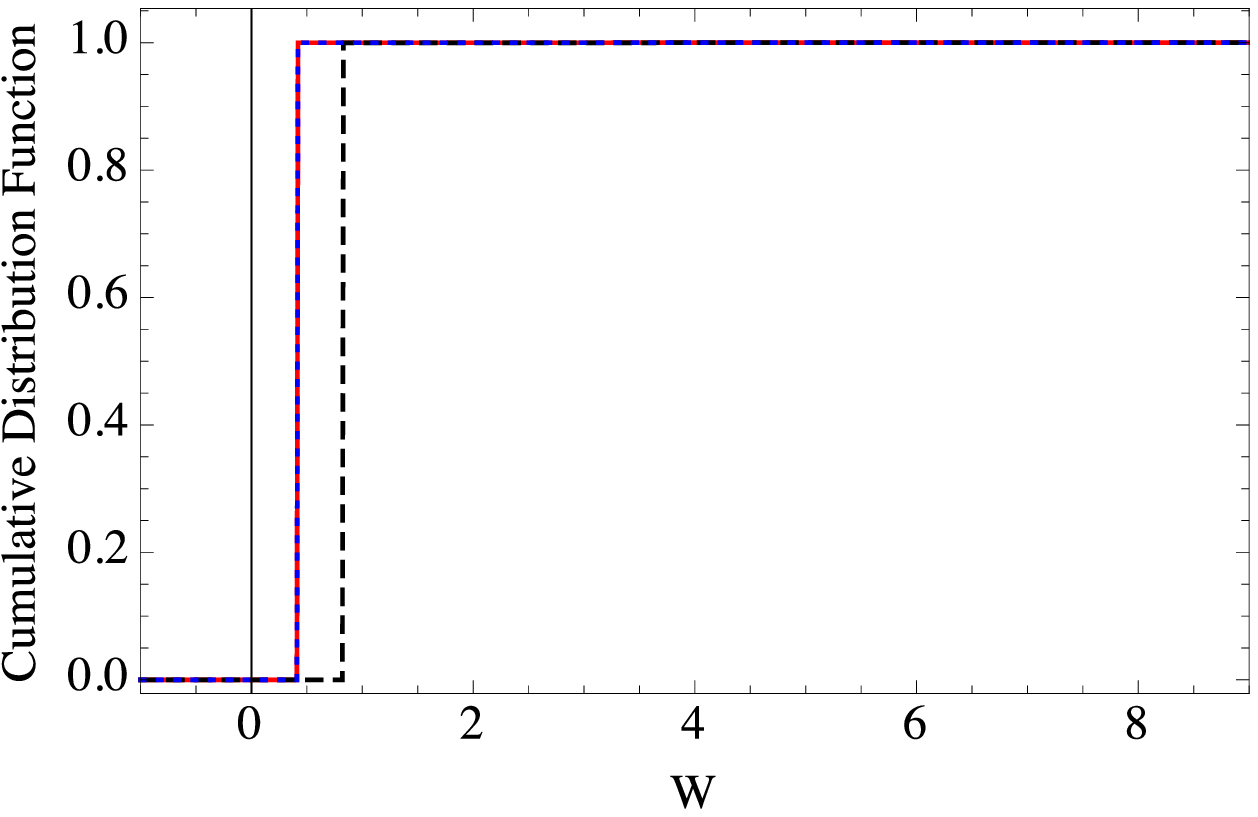}
      }
      \subfigure{%
         \label{fig5b}%
        \includegraphics[width=.42\textwidth]{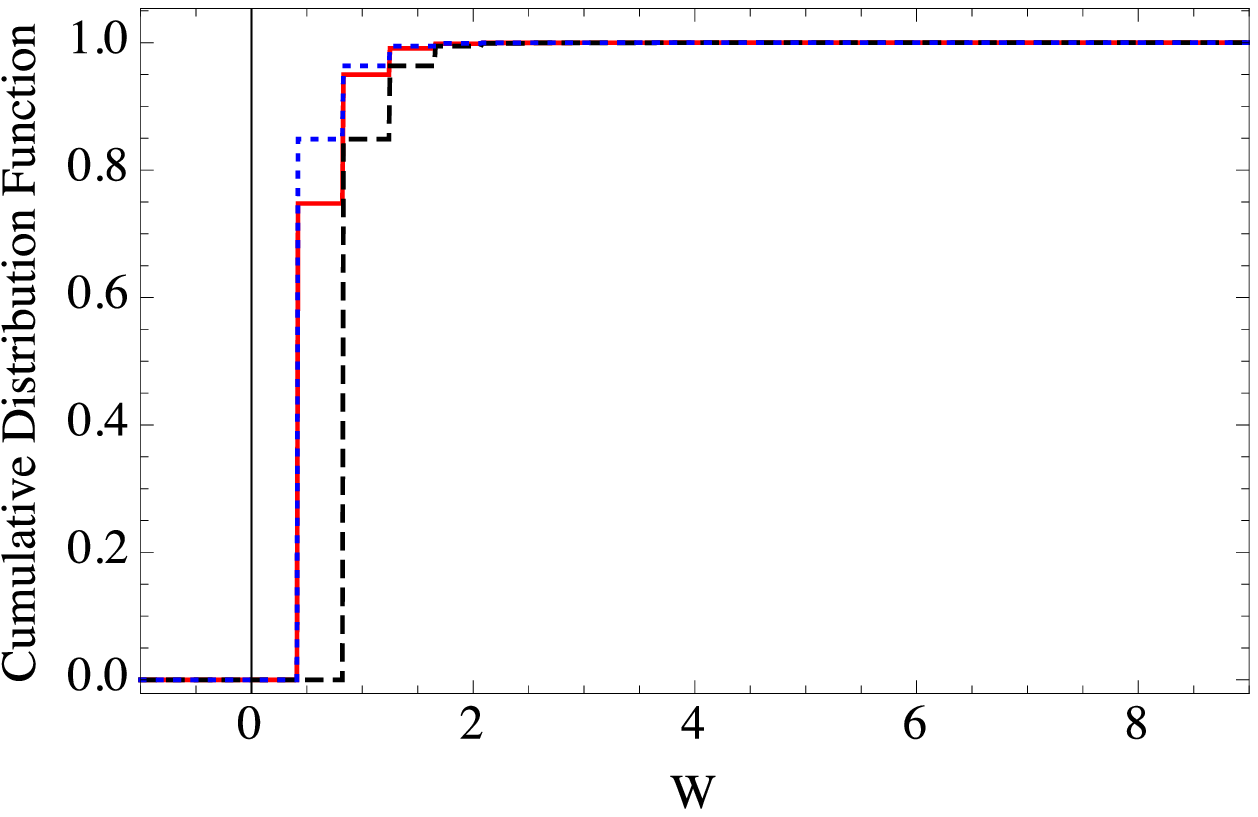}
      }
      \subfigure{%
         \label{fig5c}%
        \includegraphics[width=.42\textwidth]{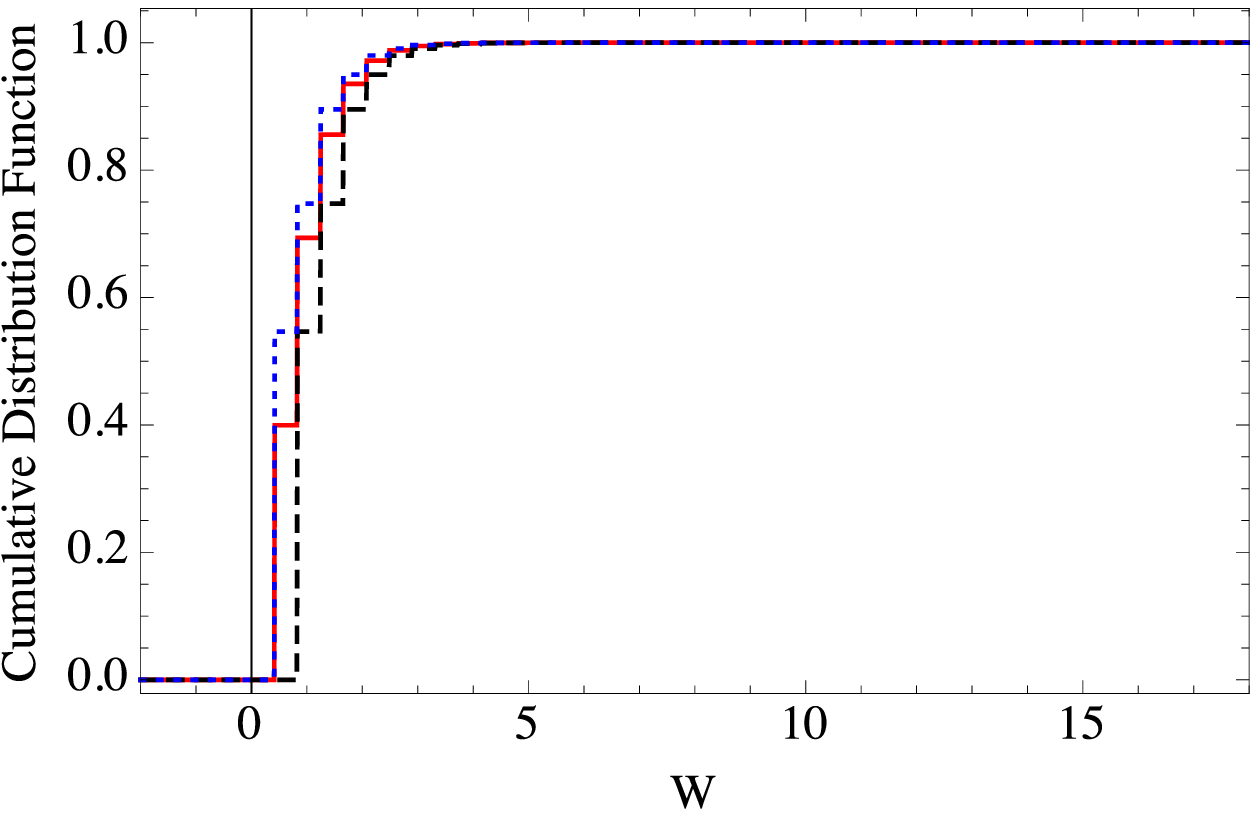}
      }
        \subfigure{%
         \label{fig5d}%
        \includegraphics[width=.42\textwidth]{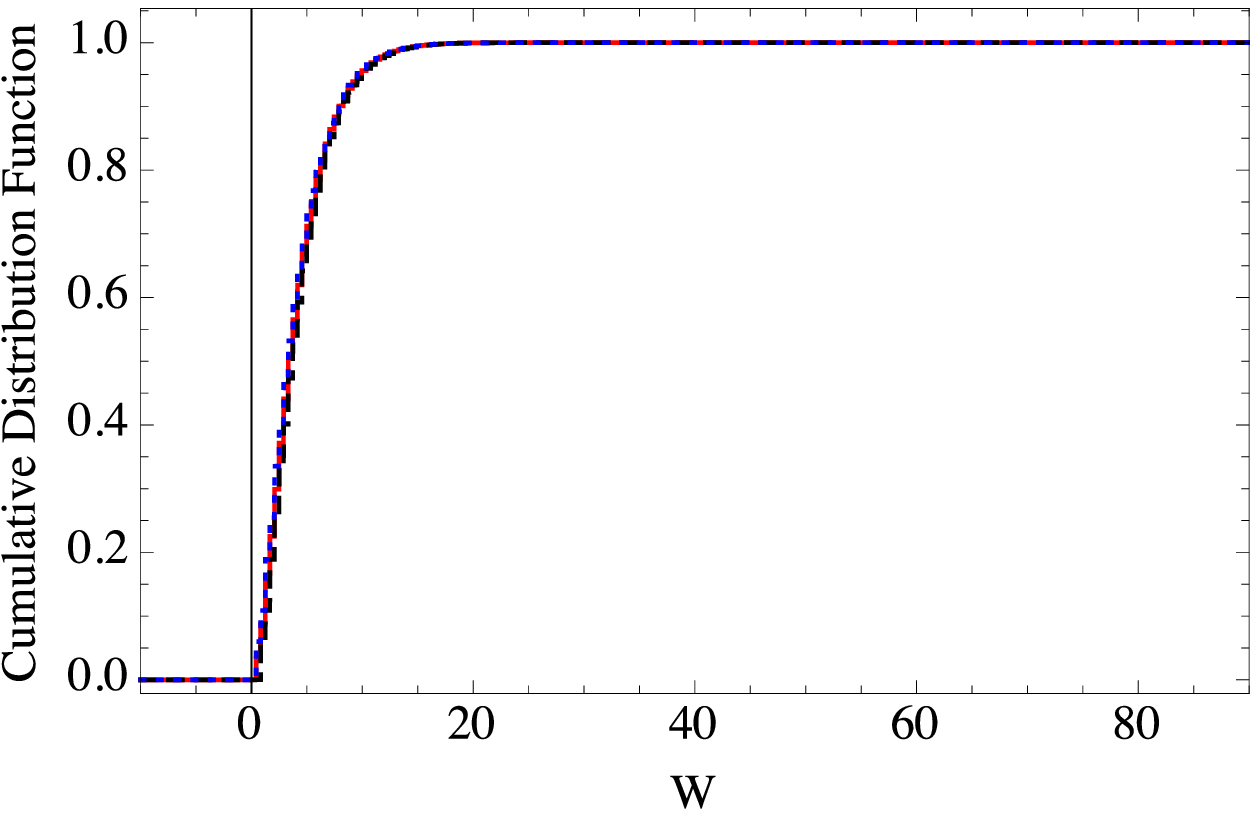}
	  }
   \caption{\label{fig:harm_2_slow}(color online) Cumulative work distribution \eqref{eq:cum} for  two  Bosons (blue, dotted line), Fermions (black, dashed line) and distinguishable particles (red, solid line) in a quenched harmonic potential \eqref{eq:harm} with $\omega_{0}=1$, $\omega_{\tau}=\sqrt{2}$, and $1/\tau=0.1$. Temperatures are from top to bottom $\beta^{-1}=0$, $\beta^{-1}=0.5$, $\beta^{-1}=1$, and $\beta^{-1}=5$. }
\end{figure}
\begin{figure}[]%
      \subfigure{%
         \label{fig6a}%
        \includegraphics[width=.42\textwidth]{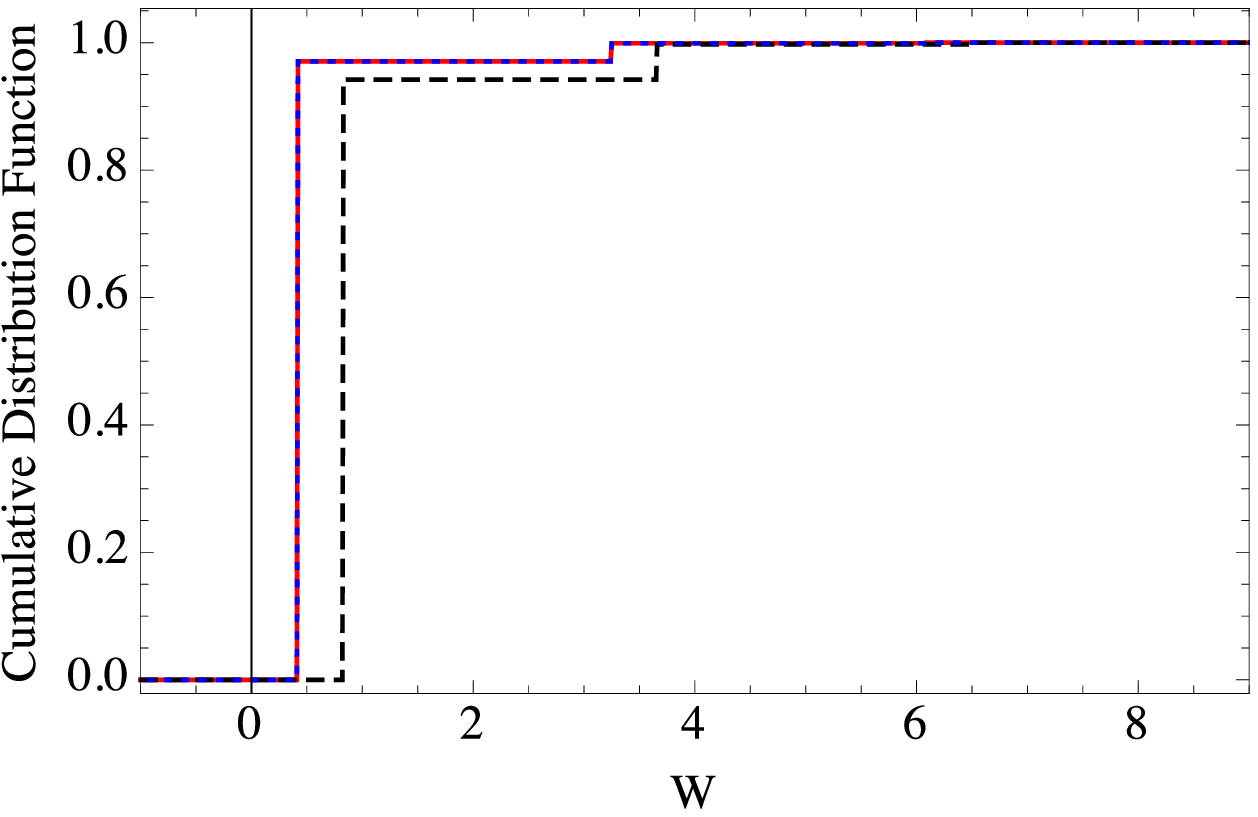}
      }
      \subfigure{%
         \label{fig6b}%
        \includegraphics[width=.42\textwidth]{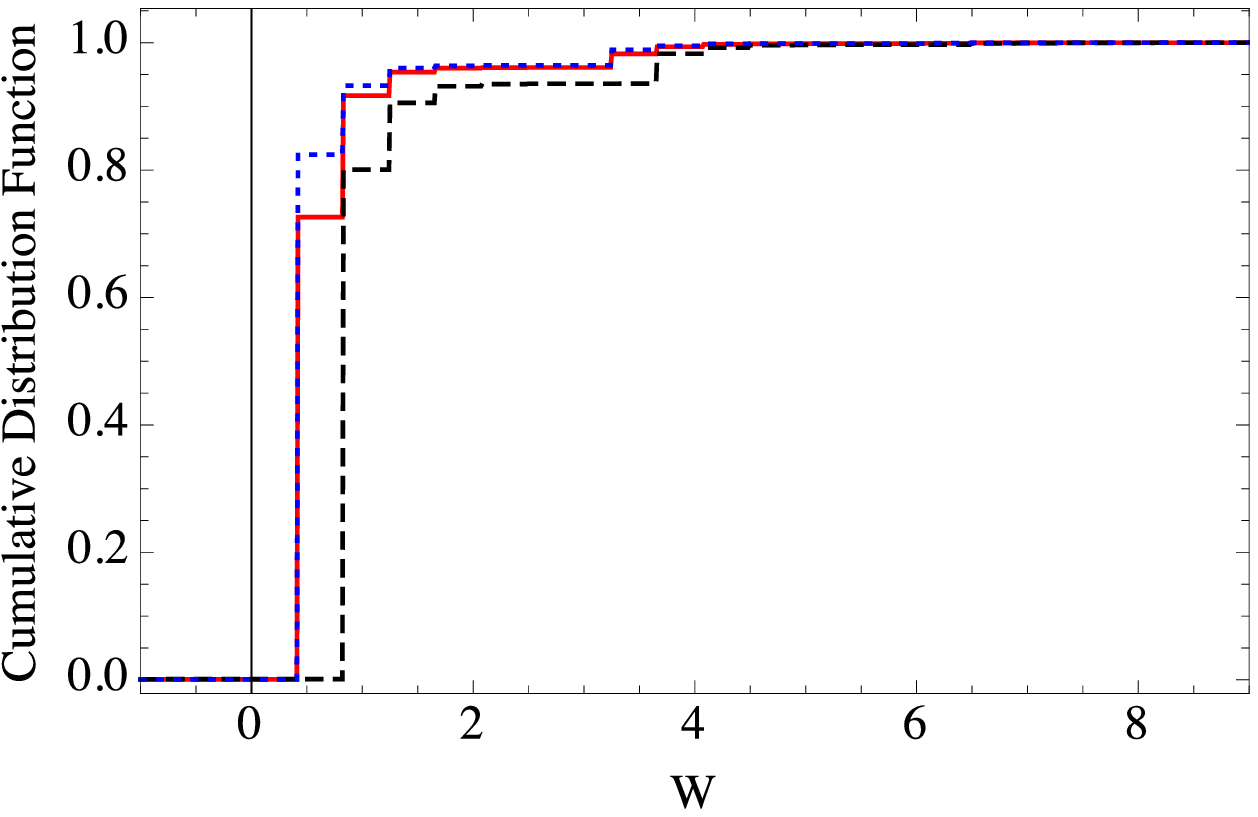}
      }
      \subfigure{%
         \label{fig6c}%
        \includegraphics[width=.42\textwidth]{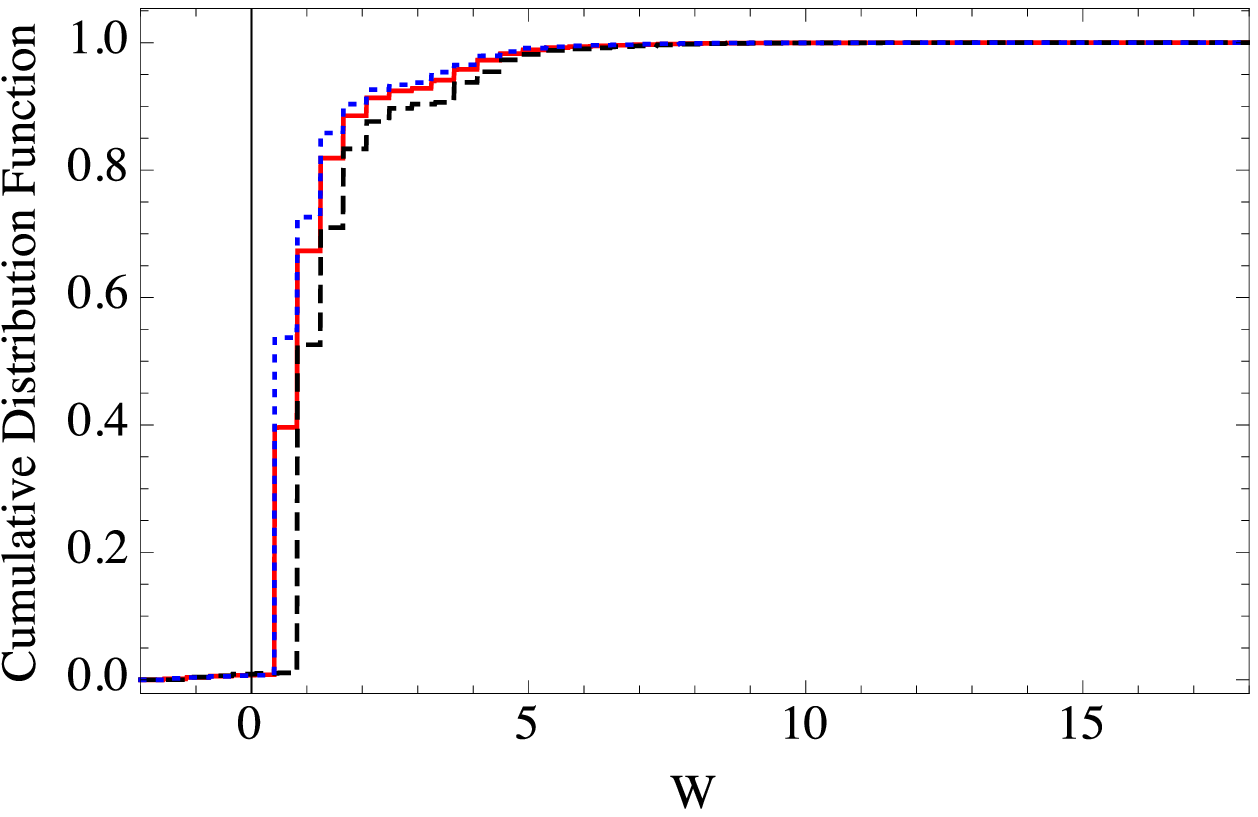}
      }
   \caption{\label{fig:harm_2_fast}(color online) Cumulative work distribution \eqref{eq:cum} for  two  Bosons (blue, dotted line), Fermions (black, dashed line) and distinguishable particles (red, solid line) in a quenched harmonic potential \eqref{eq:harm} with $\omega_{0}=1$, $\omega_{\tau}=\sqrt{2}$, and $1/\tau=100$. Temperatures are from top to bottom $\beta^{-1}=0$, $\beta^{-1}=0.5$, and $\beta^{-1}=1$. }
\end{figure}
\begin{figure}[]%
      \subfigure{%
         \label{fig7a}%
        \includegraphics[width=.42\textwidth]{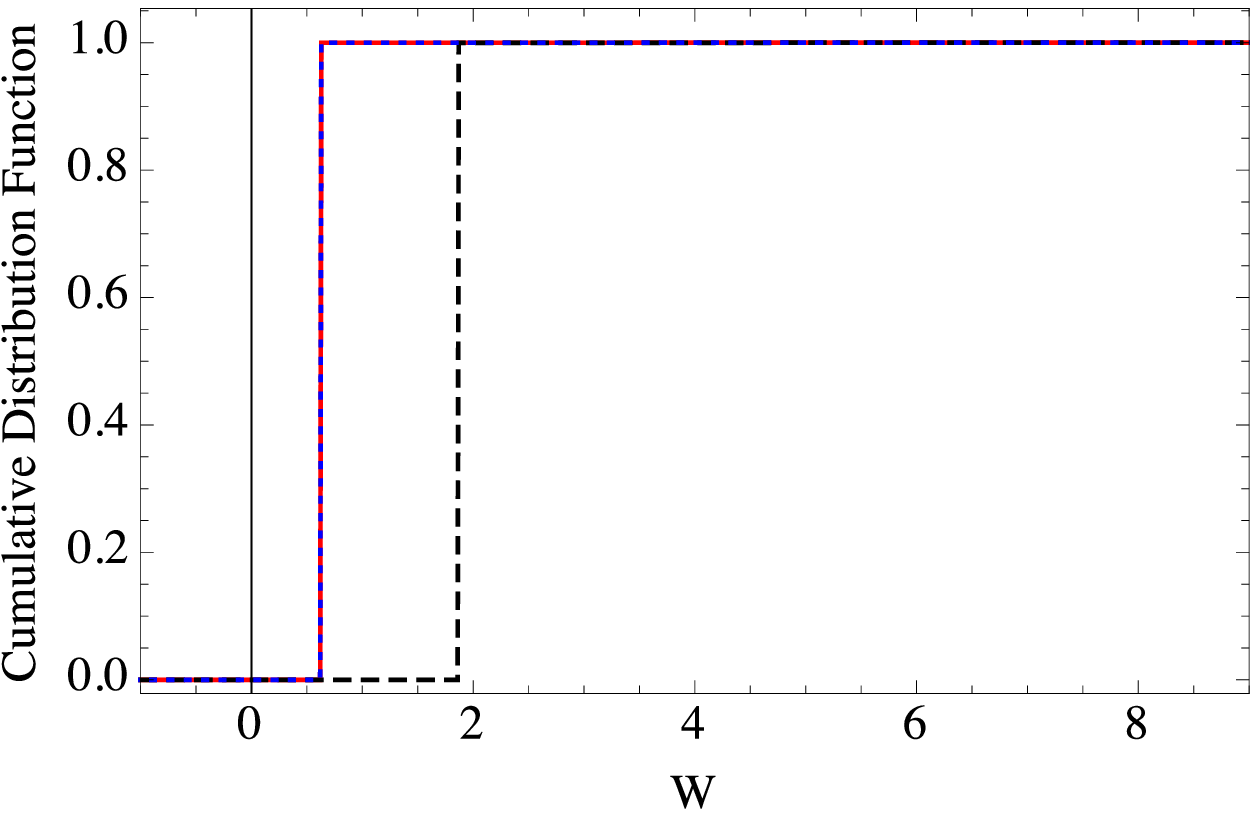}
      }
      \subfigure{%
         \label{fig7c}%
        \includegraphics[width=.42\textwidth]{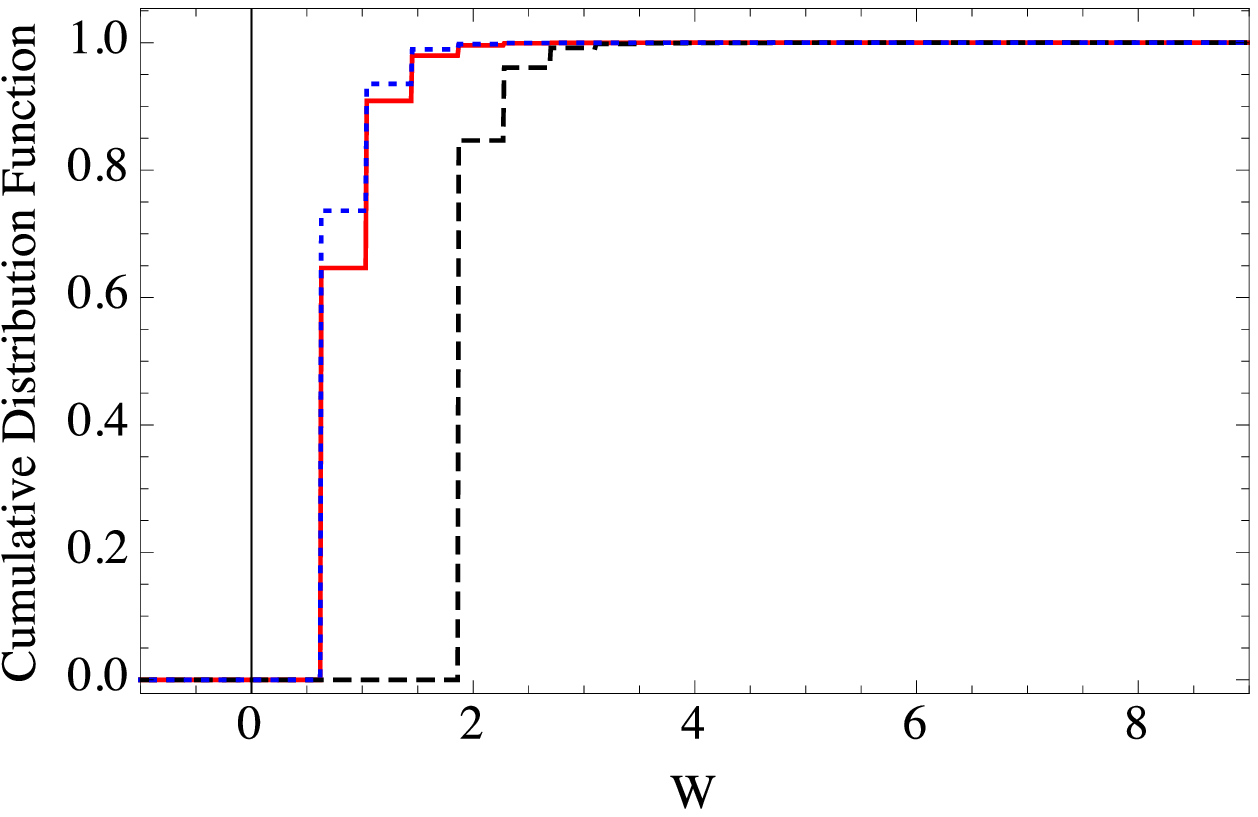}
      }
   \caption{\label{fig:harm_3_slow}(color online) Cumulative work distribution \eqref{eq:cum} for  three  Bosons (blue, dotted line), Fermions (black, dashed line) and distinguishable particles (red, solid line) in a quenched harmonic potential \eqref{eq:harm} with $\omega_{0}=1$,  $\omega_{\tau}=\sqrt{2}$, and $1/\tau=0.1$. Temperatures are from top to bottom $\beta^{-1}=0$ and $\beta^{-1}=0.5$.  }
\end{figure}
\begin{figure}[]%
      \subfigure{%
         \label{fig8a}%
        \includegraphics[width=.42\textwidth]{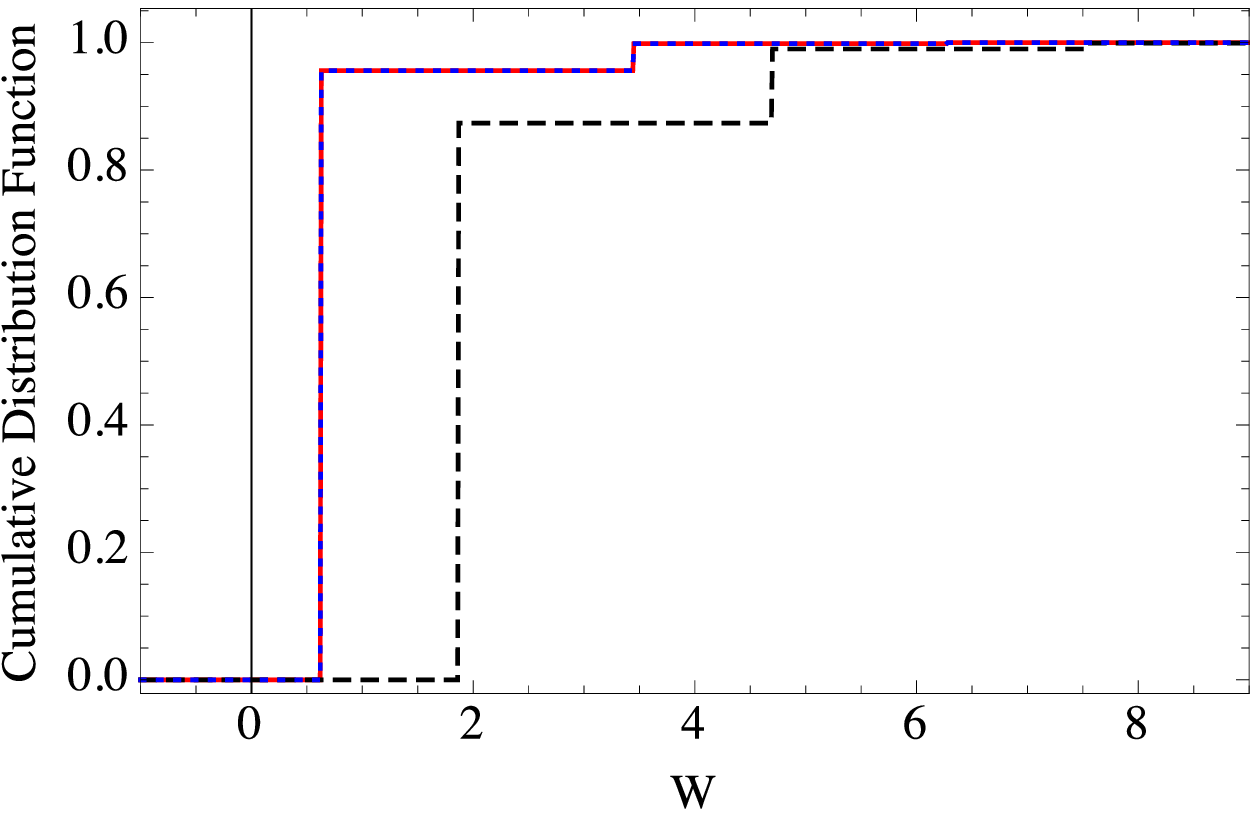}
      }
      \subfigure{%
         \label{fig8b}%
        \includegraphics[width=.42\textwidth]{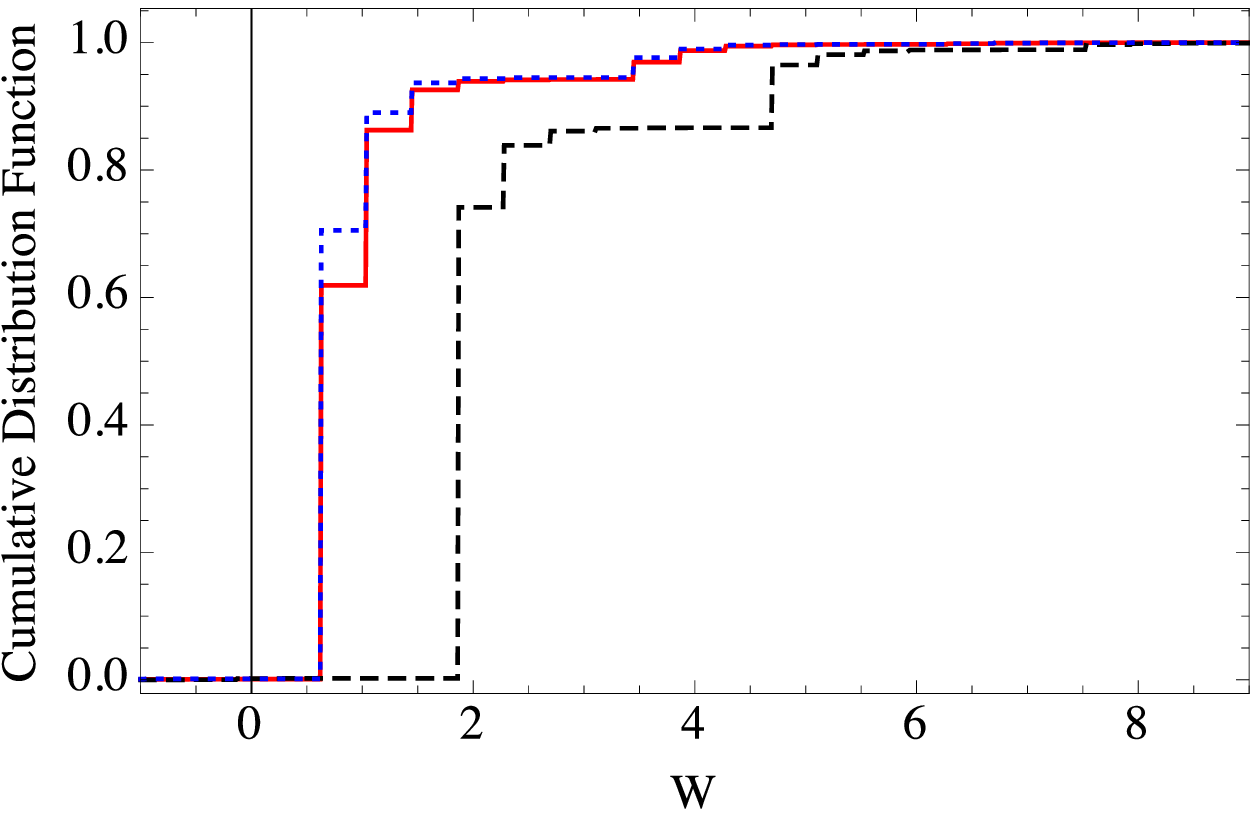}
      }
   \caption{\label{fig:harm_3_fast}(color online) Cumulative work distribution \eqref{eq:cum} for  three  Bosons (blue, dotted line), Fermions (black, dashed line) and distinguishable particles (red, solid line) in a quenched harmonic potential \eqref{eq:harm} with $\omega_{0}=1$, $\omega_{\tau}=\sqrt{2}$, and $1/\tau=100$. Temperatures are from top to bottom $\beta^{-1}=0$ and $\beta^{-1}=0.5$. }
\end{figure}

\section{\label{sec:conv}Quantum work at high temperature}

At low temperatures the thermal distributions for Bosons and for Fermions significantly differ. This is due to ``static interference'' expressed by the fact that the many-particle eigenstates can be expressed in terms of a determinant (for Fermions) and a permanent (for Bosons) in the second quantization formalism \cite{Schiff1968}. Note that the Pauli exclusion principle states that two identical Fermions cannot occupy the same single-particle state, while for Bosons there is no such a constraint. At high temperatures, however, the thermal states for Bosons and for Fermions become indistinguishable, which can be interpreted as a consequence of the static ``correspondence principle" in quantum statistical sense for $\beta \to 0$.

As we discussed in Sec.~\ref{sec:work} the transition probabilities between many-particle eigenstates for Bosons and for Fermions (\ref{transitonprobability}) can also be expressed in terms of determinants (for Fermions) and permanents (for Bosons) \cite{Scheel2004,Tichy2012,Tichy2014}. This effect can be interpreted as a ``dynamic effect'' of interference, which is independent of the initial temperature. It is thus neither obvious nor ad hoc clear whether the work distribution functions for many Bosons will converge towards that of many Fermions at high temperatures.

In the previous section, we discussed the numerical results for the quantum work distribution in two simple model systems. Our numerical results strongly suggest that at high temperatures the work distribution functions for both Bosons and Fermions do converge to that of distinguishable particles. In the following, we propose semi-heuristic arguments to show that this numerical evidence holds true for arbitrary potentials. To this end, we will make use of the representation of the work distribution in terms of its characteristic function \cite{Talkner2007}.

The characteristic function of the work distribution for a many-Boson system $G_{B}(\mu)$ and a many-Fermion system $G_{F}(\mu)$ can be expressed as \cite{Talkner2007}
\begin{equation}
G_{B/F}(\mu)=\ptr{\mathcal{H}_{B/F}^{\lambda_{0}}}{\mathcal{D}_{B/F}^{\lambda_{0}}\e{i\mu H_{H}^{\lambda_{\tau}}} \e{-i\mu H^{\lambda_{0}}}},
\label{characteristic}
\end{equation}
where
\begin{equation}
\mathcal{D}_{B/F}^{\lambda_{0}}=\frac{\e{-\beta H^{\lambda_{0}}}}{Z_{B/F}^{\lambda_{0}}}
\end{equation}
is the initial thermal equilibrium distribution with the partition function
\begin{equation}
 Z_{B/F}^{\lambda_{0}}=\ptr{\mathcal{H}_{B/F}^{\lambda_{0}}}{ \e{-\beta H^{\lambda_{0}}}}\,.
\end{equation}
Furthermore, $\mathcal{H}_{B/F}^{\lambda_{0}}$ describes the Hilbert space of the many-Boson/Fermion system with the work parameter equal to $\lambda_{0}$; $H_{H}^{\lambda_{\tau}}$ describes the Hamiltonian of the many-particle system in Heisenberg picture with the work parameter equal to $\lambda_{\tau}$ and $H^{\lambda_{0}}$ is the Hamiltonian of the many-particle system in Schr\"odinger picture with the work parameter equal to $\lambda_{0}$.

After a straightforward calculation the characteristic function of the work distribution for a system consisting of many Bosons or many Fermions can be  expressed as the characteristic functions of a single particle system (see Appendix~\ref{sec:appA}) \cite{Miller1972}
\begin{widetext}
\begin{equation}
G_{B/F}(\mu)=\frac{\sum_{(1^{\nu_{1}},2^{\nu_{2}},\cdots, N^{\nu_{N}} ) } \frac{N!}{1^{\nu_{1}}2^{\nu_{2}}\cdots N^{\nu_{N}}\nu_{1}! \nu_{2}! \cdots \nu_{N}! } \prod_{k=1}^{N}\left[ \pm \ptr{\mathcal{H}_{s}^{\lambda_{0}}}{\left[ \mathcal{G}^{\lambda_{0}} (\mu)\e{-\beta H_{s}^{\lambda_{0}}} \right]^{k}}\right]^{\nu_{k}}    }                 {\sum_{(1^{\nu_{1}},2^{\nu_{2}},\cdots, N^{\nu_{N}} ) } \frac{N!}{1^{\nu_{1}}2^{\nu_{2}}\cdots N^{\nu_{N}}\nu_{1}! \nu_{2}! \cdots \nu_{N}! } \prod_{k=1}^{N}\left[ \pm \ptr{\mathcal{H}_{s}^{\lambda_{0}}}{\e{-k\beta H_{s}^{\lambda_{0}}}} \right]^{\nu_{k}}  },
\label{derivation}
\end{equation}
\end{widetext}
Here
\begin{equation}
\mathcal{G}^{\lambda_{0}}(\mu)=\e{i\mu H_{H,s}^{\lambda_{\tau}}} \e{-i\mu H^{\lambda_{0}}_{s}},
\end{equation}
and $H_{H,s}^{\lambda_{\tau}}$ represents the Hamiltonian for a single particle in Heisenberg's picture; analogously $H^{\lambda_{0}}_{s}$ is the Hamiltonian for a single particle in Schr\"odinger's picture; $\mathcal{H}_{s}^{\lambda_{0}}$ denotes the Hilbert space of a single particle system when the work parameter is equal to $\lambda_{0}$; $(1^{\nu_{1}},2^{\nu_{2}},\cdots, N^{\nu_{N}} )$ describes a cycle notation, which corresponds uniquely to a permutation \cite{Miller1972} ($k^{\nu_{k}}$ means that there are $\nu_{k}$ $k$-cycles, $\nu_{k}\ge 0$, and $\sum_{k=1}^{N}k \times \nu_{k}=N$. The definition of $k$-cycles can be found in section 1.1 of Ref.~\cite{Miller1972}).

Equation~(\ref{derivation}) constitutes one of our main results. It is further tested and verified for the ideal quantum gas in Appendix~\ref{sec:appB}.


For convenience, we rewrite Eq.~(\ref{derivation}) in the following form as a weighted average of some characteristic function-like terms
\begin{widetext}
\begin{equation}
G_{B/F}(\mu)= \frac{\sum_{(1^{\nu_{1}},2^{\nu_{2}},\cdots, N^{\nu_{N}} ) } \mathcal{M}_{(1^{\nu_{1}},2^{\nu_{2}},\cdots, N^{\nu_{N}} )}^{B/F} R_{(1^{\nu_{1}},2^{\nu_{2}},\cdots, N^{\nu_{N}} )}(\beta) G_{(1^{\nu_{1}},2^{\nu_{2}},\cdots, N^{\nu_{N}} )}(\mu) }                 {\sum_{(1^{\nu_{1}},2^{\nu_{2}},\cdots, N^{\nu_{N}} ) } \mathcal{M}_{(1^{\nu_{1}},2^{\nu_{2}},\cdots, N^{\nu_{N}} )}^{B/F} R_{(1^{\nu_{1}},2^{\nu_{2}},\cdots, N^{\nu_{N}} )}(\beta)  },
\label{rewritten}
\end{equation}
\end{widetext}
where $\mathcal{M}_{(1^{\nu_{1}},2^{\nu_{2}},\cdots, N^{\nu_{N}} )}^{B/F}$ are temperature-independent values (see Appendix~\ref{sec:appA} for details) and
\begin{equation}
\begin{split}
&R_{(1^{\nu_{1}},2^{\nu_{2}},\cdots, N^{\nu_{N}} )}(\beta)\\
&=\frac{\prod_{k=1}^{N}\left[  \ptr{\mathcal{H}_{s}^{\lambda_{0}}} { \e{-k\beta H_{s}^{\lambda_{0}}}} \right]^{\nu_{k}}}{\left[ \ptr{\mathcal{H}_{s}^{\lambda_{0}}}{\e{-\beta H_{s}^{\lambda_{0}}}}\right]^{N}},
\end{split}
\end{equation}
\begin{equation}
\begin{split}
&G_{(1^{\nu_{1}},2^{\nu_{2}},\cdots, N^{\nu_{N}} )}(\mu)\\
&=\prod_{k=1}^{N}\left[\frac{\ptr{\mathcal{H}_{s}^{\lambda_{0}}} {\left[ \mathcal{G}^{\lambda_{0}} (\mu)\e{-\beta H_{s}^{\lambda_{0}}} \right]^{k}}}{  \ptr{\mathcal{H}_{s}^{\lambda_{0}}} { \e{-k\beta H_{s}^{\lambda_{0}}}}}\right]^{\nu_{k}}.
\end{split}
\end{equation}
In the high temperature limit, the dominant contribution in both the numerator and the denominator of $G_{B/F}(\mu)$ (see Eq.~(\ref{rewritten})) stems from the trivial identity permutation, which is $(1^{N},2^{0},\cdots, N^{0} )$ in cycle notation. If we keep only the leading term in both the denominator and the numerator, the characteristic function (\ref{rewritten}) can be simplified to read  (for details please see Appendix \ref{sec:appC})
\begin{equation}
G_{B/F}(\mu)\approx \left[\frac{\ptr{\mathcal{H}_{s}^{\lambda_{0}}}{\mathcal{G}^{\lambda_{0}} (\mu)\e{-\beta H_{s}^{\lambda_{0}}}}}{\ptr{\mathcal{H}_{s}^{\lambda_{0}}}{\e{-\beta H_{s}^{\lambda_{0}}}}} \right]^{N}.
\label{asymptotic}
\end{equation}
The latter expression is exactly the characteristic function of the work distribution for distinguishable particles. Thus, we have demonstrated that, in the high temperature limit, the characteristic function of work distribution functions for both Bosons and Fermions converge towards that of distinguishable particles, and hence to each other. Since there is a one-to-one map between the work distribution and its corresponding characteristic function \cite{Talkner2007}, we conclude that in the high temperature limit, the work distributions for Bosons and Fermions converge.

This conclusion can be understood intuitively: Firstly, in the limit of high temperatures, for most of the many-particle eigenstates, usually a single-particle state is occupied by at most one particle, and the occupations of Bosons, Fermions and distinguishable particles in the single-particle states become similar. Accordingly, the distributions of the multi-particle eigenenergies (\ref{probability}) for three kinds of particles become similar. This is the so-called ``static" correspondence principle in the quantum statistical sense. Secondly, for a given finite-speed work protocol, if the average velocity of the particles in a high-lying multi-particle eigenstate is much higher than the speed of varying the parameter, the process can be roughly regarded as a ``quasistatic" process. In the thermal state of infinite temperature, all energy eigenstates have equal probability. That means for a given finite-speed protocol, for most of the initial eigenstates (sampled from the thermal state at the infinite temperature), the processes can be regarded as quasistatic processes. Thirdly, if the dynamic process is a quasistatic process, the work value usually can be assumed to be proportional to the initial eigenenergy (at least it is true for the 1D piston and the 1D harmonic oscillator (\ref{proportionality})), and thus the work distributions will share the same properties as the distributions of the multi-particle eigenenergies (\ref{probability}). Based on these observations, one can infer that for an arbitrary finite-speed protocol and in the limit of infinite temperature the work distribution functions for three kinds of particles converge.

It is worth emphasizing that this result does not depend on the specifics of the model, and, hence, holds for any system of many noninteracting, identical particles.

\section{\label{sec:con}Discussion and conclusion}

In this paper we have studied the effect of indistinguishability (quantum interference of identical, noninteracting particles) on the quantum work distribution. We have found that the work distribution can be computed from the time evolution matrix for single particles. Then, the transition amplitudes between multi-particle states are given by the Slater-determinant (for Fermions) or the permanent (for Bosons).

\subsection{Determinants and permanents}

Generally, the computation of the permanent of a matrix is much more involved than the computation of the determinant -- despite the apparent similarity of the definitions \cite{Tichy2014}: In particular, the determinant obeys several algebraic rules and symmetries, e.g., the product rule $\det{A B}=\det{A} \det{B}$ and the invariance under unitary transformation, which allow the determinant to be  evaluated in polynomial time. For a $N\times N$ matrix, e.g., the elementary Gaussian algorithm needs $O(N^{3})$ operations \cite{Tichy2014}. Although the permanent has a similar structure the omission of the alternating sign makes all the difference, and all known strategies for an efficient evaluation of the determinant fail for the permanent. In general, a permanent can only be computed in exponential time, even when applying Ryser's algorithm \cite{Tichy2014} -- the most efficient algorithm known to date. Therefore, the development of both exact and approximate algorithms for computing the permanent of a matrix is an active area of research.

For our problem this means that the computation of the work density for Fermions is much more feasible than for Bosons. The exponential increase of the complexity of computing the permanent limits the study of the work distribution to at most 25 Bosons \cite{Tichy2014}.

This restriction might be lifted by a novel development in the field of quantum information known as Boson Sampling -- a shortcut to quantum computing \cite{Aaronson2013}. In this technique the bosonic distribution is obtained from interfering photons in a random optical network. However, practical applications of Boson Sampling are still under active research \cite{Tillmann2013}.

\subsection{Quasistatic limit}

Our expression for the characteristic function (\ref{derivation}) is valid for both quasistatic and nonquasistatic processes.  To the best of our knowledge, previous studies have been restricted to classical distinguishable particles \cite{Lua2005}, quasistatic changes \cite{Crooks2007}, and the infinite piston system \cite{Bena2005}.

For the sake of completeness we, thus, briefly show how the expression for quasistatic processes \cite{Crooks2007} can be obtained from our general formula \eqref{derivation}. For very slow driving one can assume that the energy levels remain almost constant. Thus the characteristic function for Bosons/Fermions, $G_{B/F}(\mu)$, can be written as
\begin{widetext}
\begin{equation}
G_{B/F}(\mu)=\frac{\sum_{(1^{\nu_{1}},2^{\nu_{2}},\cdots, N^{\nu_{N}} ) } \frac{N!}{1^{\nu_{1}}2^{\nu_{2}}\cdots N^{\nu_{N}}\nu_{1}! \nu_{2}! \cdots \nu_{N}! } \prod_{k=1}^{N}\left[ \pm \sum_{i=1}^{\infty} \e{ik\mu E_{i}^{\lambda_{\tau}}} \e{-k(i\mu+\beta) E_{i}^{\lambda_{0}}} \right]^{\nu_{k}}    }                 {\sum_{(1^{\nu_{1}},2^{\nu_{2}},\cdots, N^{\nu_{N}} ) } \frac{N!}{1^{\nu_{1}}2^{\nu_{2}}\cdots N^{\nu_{N}}\nu_{1}! \nu_{2}! \cdots \nu_{N}! } \prod_{k=1}^{N}\left[ \pm \sum_{i=1}^{\infty} \e{-k\beta E_{i}^{\lambda_{0}}} \right]^{\nu_{k}}  },
\label{quasistaticprocess}
\end{equation}
\end{widetext}
and analogously for distinguishable particles
\begin{equation}
G(\mu)=\left[\frac{\sum_{i=1}^{\infty} \e{i\mu E_{i}^{\lambda_{\tau}}} \e{-(i\mu+\beta) E_{i}^{\lambda_{0}}}}{\sum_{i=1}^{\infty} \e{-\beta E_{i}^{\lambda_{0}}}} \right]^{N}.
\label{diagonal}
\end{equation}
We further assume the quasistatic work to be proportional to the initial eigenenergies
\begin{equation}
W=E_{i}^{\lambda_{\tau}}-E_{i}^{\lambda_{0}}=\alpha\, E_{i}^{\lambda_{0}}, \quad \forall i,
\label{proportionality}
\end{equation}
which is justified, for instance, for a particle in a 1D piston or in a 1D harmonic potential. For this kind of systems we also can assume that each eigenenergy can be written as a power of the quantum number, and we have
\begin{equation}
E_{i}^{\lambda_{0}} =E_{0}\times i^{p}\,.
\end{equation}
Under these assumptions the numerator of the characteristic function for a single particle can be approximately expressed as
\begin{equation}
\begin{split}
&\quad\sum_{i=1}^{\infty} \e{i\mu \alpha E_{i}^{\lambda_{0}}} \e{-\beta E_{i}^{\lambda_{0}}}\approx\\
&\int\limits_{-\infty}^{\infty}\td W\, \frac{\e{-\beta W/\alpha} \e{i \mu W}}{p|W|} \left(\frac{W}{\alpha E_{0}} \right)^{1/p}\Theta(\alpha W),
\end{split}
\end{equation}
where $\Theta(\cdot)$ is the Heaviside step function.

Specifically, if the system is a particle in a 1D piston, the quasistatic work distribution becomes
\begin{equation}
\begin{split}
&\mc{P}_{s}(W)=\frac{\frac{\e{-\beta W/\alpha}}{2|W|} \left(\frac{W}{\alpha E_{0}} \right)^{1/2}\Theta(\alpha W)}{\int\limits_{-\infty}^{\infty}\td W\,\frac{\e{-\beta W/\alpha}}{2|W|} \left(\frac{W}{\alpha E_{0}} \right)^{1/2}\Theta(\alpha W)}\\
&=\frac{\beta}{|\alpha|\Gamma(1/2)} \left(\frac{\beta W}{\alpha} \right)^{\frac{1}{2}-1} \e{-\beta W/\alpha}\, \Theta(\alpha W).
\end{split}
\end{equation}
For $N$ distinguishable particles, the work distribution function can be obtained by replacing the $1/2$ with $N/2$ and due to the additivity of the Gamma distribution we obtain
\begin{equation}
\mc{P}(W)=\frac{\beta}{|\alpha|\Gamma(N/2)} \left(\frac{\beta W}{\alpha} \right)^{\frac{N}{2}-1} \e{-\beta W/\alpha}\, \Theta(\alpha W).
\label{workdistributionprime}
\end{equation}
The latter result coincides with the expression for quasistatically compressing (or expanding) $N$-particle gas inside a piston, which was previously derived in Ref.~\cite{Crooks2007}.

The discussion of the convergence in Sec. ~\ref{sec:conv} can be generalized to the high temperature limit for any given protocol. In particular, in the high temperature limit, for an arbitrary finite-speed protocol, most initial preparations of the system will undergo approximately quasistatic evolution, which can be recognized as a special realization of the quasistatic limit.

\subsection{Fluctuation theorems}

We have seen in the above discussion that quantum interference effects the structure of the work distribution. It is worth emphasizing, however, that the quantum Jarzynski equality \cite{Jarzynski1997a,Campisi2011} and the quantum Crooks fluctuation theorem \cite{Crooks1999,Campisi2011,Quan2008a} remain valid as the validity of these two nonequilibrium work relations does not depend on the details of the model or the quench protocol, cf. Appendix \ref{sec:appB}. Also note that even if the initial $N$-particle states occupies many different single particle states (as is allowed for both Bosons and Fermions) the average work will be identical, and only the distribution differs. This is due to the particles being noninteracting but interfering \cite{Tichy2014}.

\subsection{Concluding remarks}

In summary, we have studied the interference of identical particles and its influence on the quantum work distribution function in nonequilibrium processes. To this end, we have applied the results for the transition amplitudes between many-particle eingenstates \cite{Tichy2014}. At low temperatures, the work distributions for Bosons and for Fermions significantly differ due to the interference of identical particles, and the larger the particle number, the more prominent is the distinction between the work distributions for Bosons and for Fermions. In principle, the work distribution function for many Bosons or many Fermions can be evaluated as long as the transition amplitudes between single particle eigenstates can be calculated. In practice, however, the work distribution function for Bosons is computationally much more involved than Fermions due to the complexity of the computation of the permanent of a matrix.

As a case study we have numerically calculated the work distribution function for two and three identical particles in the 1D piston and 1D harmonic potential, and have demonstrated our theoretical findings.

In the limit of high temperature, the work distribution functions for Bosons and Fermions converge, and we have given a heuristic analysis for this observation by utilizing the representation theory of the symmetric group as well as a qualitative argument based on the static ``correspondence principle" and the quantum adiabaticity. Therefore, our study suggests a dynamic ``correspondence principle" of work distribution functions in the quantum statistical sense.

\acknowledgments
HTQ gratefully acknowledges support from the National Science Foundation of China under grant 11375012, and The Recruitment Program of Global Youth Experts of China. SD acknowledges financial support by the National Science Foundation (USA) under grant DMR-1206971, and by the U.S. Department of Energy through a LANL Director's Funded Fellowship.

\appendix


\section{\label{sec:appD} single particle transition matrix for the harmonic oscillator}

The parametric harmonic oscillator with Hamiltonian \eqref{eq:harm} has been extensively studied in the literature. For the sake of completeness we summarize in this appendix several expressions that were used to compute the plots in Fig.~\ref{fig:harm_2_slow}-\ref{fig:harm_3_fast}, and we also a correct minor typographical error that appeared in a previous publication \cite{Deffner2010}.

The single particle transition matrix has been derived in Ref.~\cite{Deffner2010}
\begin{equation}
\begin{split}
U_{m,n}^\tau=&\sqrt[4]{\omega_0\omega_{\tau}} \sqrt{\frac{n!\,m!\,\zeta^n {\zeta^*}^m }{2^{n+m-1}i\sigma^{n+m+1}}}\\
\times&\sum\limits_{l=0}^{\min{(m,n)}}\frac{[-2 i \sqrt{2/(Q^*-1)}]^l}{l!\,[(n-l)/2]!\,[(m-l)/2]!} \ .
\end{split}
\end{equation}
According to the selection~rule $m=n\pm2k$, $l$ runs over even numbers only, if $m$, $n$ are even, and over odd numbers only, if $m$, $n$ are odd.

The explicit expression for the matrix elements  $U_{m,n}^\tau$ then reads for even elements
\begin{widetext}
\begin{equation}
\begin{split}
 U_{2\mu,2\nu}^\tau=\sqrt{\frac{2\nu!2\mu!}{2^{2\nu+2\mu-1}i}}\sqrt{\frac{\zeta^{2\nu} {\zeta^*}^{2\mu}}{\sigma^{2\nu+2\mu+1}}}\frac{\sqrt[4]{\omega_0\omega_{\tau}}}{\Gamma(\mu+1)\,\Gamma(\nu+1)} {_2 F_1}\left(-\mu,\,-\nu;\,\frac{1}{2};\,\frac{2}{1-Q^*}\right)
\end{split}
\end{equation}
and for odd elements
\begin{equation}
\begin{split}
 U_{2\mu+1,2\nu+1}^\tau=-\sqrt{\frac{8i\,(2\nu+1)!(2\mu+1)!}{(Q^*-1)\,2^{2\nu+2\mu+1}}}\sqrt{\frac{\zeta^{2\nu+1}{\zeta^*}^{2\mu+1}}{\sigma^{2\nu+2\mu+1}}}\frac{\sqrt[4]{\omega_0\omega_{\tau}}}{\Gamma(\mu+1)\,\Gamma(\nu+1)} {_2 F_1}\left(-\mu,\,-\nu;\,\frac{3}{2};\,\frac{2}{1-Q^*}\right).
\end{split}
\end{equation}
\end{widetext}
We have here introduced the hypergeometric function $_2 F_1$ \cite{Abramowitz1964} in order to simplify the sums and write the matrix elements $U_{m,n}^\tau$ in closed form. $\Gamma(x)$ denotes the Gamma function.

We further introduced the complex parameters,
\begin{eqnarray}
 \zeta  &=&  \omega_\tau\omega_0 X_\tau -\omega_0 i \dot X_\tau+\omega_\tau i Y_\tau +\dot Y_\tau\nonumber\\
 |\zeta|^2 &=&  2\omega_0\omega_{\tau}\,(Q^*-1) \\
 \sigma &=& \omega_\tau\omega_0 X_\tau-\omega_0 i \dot X_\tau -\omega_\tau i Y_\tau -\dot Y_\tau\nonumber\\
|\sigma|^2  &=&   2\omega_0\omega_\tau(Q^*+1)
\end{eqnarray}
with
\begin{equation}
Q^*=\frac{1}{2 \omega_0\omega_\tau}\left[\omega_0^2\left(\omega_\tau^2 X_\tau^2+\dot{X}_\tau^2\right)\left(\omega_\tau^2 Y_\tau^2+\dot{Y}_\tau^2\right)\right]
\end{equation}
and where $X_t$ and $Y_t$ are solutions of the classical, force free equation of motion,
\begin{equation}
\ddot{\xi_t}+\omega_t^2\, \xi_t=0
\end{equation}
with $X_{t=0}=0$, $\dot{X}_{t=0}=1$ and $Y_{t=0}=1$, $\dot{Y}_{t=0}=0$.

\section{\label{sec:appA}derivation of the characteristic function of work distribution functions (\ref{derivation})}

The characteristic function of the work distribution can be obtained by using the representation theory of the symmetric group (see section 4.4 of Ref.~\cite{Miller1972}).

From Eq. (\ref{characteristic}) it follows that the characteristic function for a many-particle system can be expressed as
\begin{equation}
\begin{split}
&G_{B/F}(\mu)=\\
&\frac{\ptr{\mathcal{H}_{B/F}^{\lambda_{0}}}{\e{i\mu H_{H}^{\lambda_{\tau}}} \e{-i\mu H^{\lambda_{0}}} \e{-\beta H^{\lambda_{0}}}}}{\ptr{\mathcal{H}_{B/F}^{\lambda_{0}}}{ \e{- \beta H^{\lambda_{0}}}}}.
\end{split}
\label{characteristicfuntion}
\end{equation}
Using the representation theory of the symmetric group, we further have
\begin{widetext}
\begin{equation}
\begin{split}
G_{B/F}(\mu)=&\frac{\frac{1}{N!}\sum_{\mathcal{P} \in S_{N}} (\pm)^{p(\mathcal{P})}    \ptr{\mathcal{H}^{\lambda_{0}}} {\e{i\mu H_{H}^{\lambda_{\tau}}} \e{-i\mu H^{\lambda_{0}}} \e{-\beta H^{\lambda_{0}}}\, \mathcal{P}}                  }
{\frac{1}{N!}\sum_{\mathcal{P} \in S_{N}} (\pm)^{p(\mathcal{P})}       \ptr{\mathcal{H}^{\lambda_{0}}}{\e{- \beta H^{\lambda_{0}}}\, \mathcal{P}}       }\\
=& \frac{\sum_{(1^{\nu_{1}},2^{\nu_{2}},\cdots, N^{\nu_{N}} ) } \mathcal{M}_{(1^{\nu_{1}},2^{\nu_{2}},\cdots, N^{\nu_{N}} )}^{B/F} \prod_{k=1}^{N}\left[  \ptr{\mathcal{H}_{s}^{\lambda_{0}}} {\left[ \mathcal{G}^{\lambda_{0}} (\mu)\e{-\beta H_{s}^{\lambda_{0}}} \right]^{k}} \right]^{\nu_{k}}    }                 {\sum_{(1^{\nu_{1}},2^{\nu_{2}},\cdots, N^{\nu_{N}} ) } \mathcal{M}_{(1^{\nu_{1}},2^{\nu_{2}},\cdots, N^{\nu_{N}} )}^{B/F} \prod_{k=1}^{N}\left[  \ptr{\mathcal{H}_{s}^{\lambda_{0}}} { \e{-k\beta H_{s}^{\lambda_{0}}}} \right]^{\nu_{k}}  },
\end{split}
\label{A2}
\end{equation}
\end{widetext}
where the equality sign holds only for non-interacting many-particle systems, and $S_{N}$ is a group comprised of all permutation operators on $\mathcal{H}^{\lambda_{0}}=\mathcal{H}_{s}^{\lambda_{0}} \otimes \mathcal{H}_{s}^{\lambda_{0}} \otimes \cdots \otimes \mathcal{H}_{s}^{\lambda_{0}}$. The elements of the group are denoted by $\mathcal{P}$, and $p(\mathcal{P})$ is the transposition number of $\mathcal{P}$; $\mathcal{M}_{(1^{\nu_{1}},2^{\nu_{2}},\cdots, N^{\nu_{N}} )}^{B/F}$ is the number of permutations belonging to $(1^{\nu_{1}},2^{\nu_{2}},\cdots, N^{\nu_{N}} )$ type, which has been studied in combinational mathematics and satisfies
\begin{equation}
\begin{split}
&\mathcal{M}_{(1^{\nu_{1}},2^{\nu_{2}},\cdots, N^{\nu_{N}} )}^{F}=(-)^{p(\mathcal{P})} \mathcal{M}_{(1^{\nu_{1}},2^{\nu_{2}},\cdots, N^{\nu_{N}} )}^{B}\\
&=(-)^{N-\sum_{k=1}^{N}\nu_{k}} \mathcal{M}_{(1^{\nu_{1}},2^{\nu_{2}},\cdots, N^{\nu_{N}} )}^{B},
\end{split}
\label{A3}
\end{equation}
and
\begin{equation}
\mathcal{M}_{(1^{\nu_{1}},2^{\nu_{2}},\cdots, N^{\nu_{N}} )}^{B}=\frac{N!}{1^{\nu_{1}}2^{\nu_{2}}\cdots N^{\nu_{N}}\nu_{1}! \nu_{2}! \cdots \nu_{N}!}.
\label{A4}
\end{equation}
Substituting Eqs.~(\ref{A3}) and (\ref{A4}) into Eq.~(\ref{A2}) we finally obtain the characteristic function \eqref{derivation}.

Before concluding this section, we would like to point out that the relation
\begin{equation}
\mathrm{tr}_{\mathcal{H}_{B/F}^{\lambda_{0}}} \left( \hat{A} \right)=\frac{1}{N!}\sum_{\mathcal{P} \in S_{N}} (\pm)^{p(\mathcal{P})}    \mathrm{tr}_{\mathcal{H}^{\lambda_{0}}} \left( \hat{A} \mathcal{P} \right),
\end{equation}
holds true even when the particles are interacting. Here $\hat{A}$ is an operator of multi-particle system.

\section{\label{sec:appB}derivation of the equation of state of the ideal quantum gas from the characteristic function (\ref{derivation})}
In this appendix we derive the equation of state of the ideal quantum gas inside a piston from the characteristic function (\ref{derivation}). This derivation is used as a self-consistent check to support the validity of Eq.~(\ref{derivation}). For a 1D piston system we define $\alpha^{F}=\lambda_{0}^{2}/\lambda_{\tau}^{2}-1$ and $\alpha^{R}=\lambda_{\tau}^{2}/\lambda_{0}^{2}-1$ as the ratio of the work over the initial eigenenergy of the system for the forward ($\lambda_{0} \to \lambda_{\tau}$) and the reverse ($\lambda_{\tau} \to \lambda_{0}$) process. By making use of Eqs.~(\ref{quasistaticprocess}) and following the procedure from Eq.~(\ref{diagonal}) to Eq.~(\ref{workdistributionprime}) we can obtain the work distribution for Bosons or Fermions undergoing the quasistatic process, and it can be checked that the work distribution function satisfies the Crooks Fluctuation Theorem \cite{Crooks1998,Crooks1999}
\begin{widetext}
\begin{equation}
\mc{P}_{B/F}^{F}(W)=\frac{\sum_{(1^{\nu_{1}},2^{\nu_{2}},\cdots, N^{\nu_{N}} ) } \frac{N!}{ \prod_{k=1}^{N} k^{3\nu_{k}/2}  \prod_{k=1}^{N} \nu_{k}!  } \left[ \pm  \frac{\lambda_{\tau}}{\lambda_{T}}            \right]^{\sum_{k=1}^{N}\nu_{k}}    }                 {\sum_{(1^{\nu_{1}},2^{\nu_{2}},\cdots, N^{\nu_{N}} ) } \frac{N!}{\prod_{k=1}^{N} k^{3\nu_{k}/2}  \prod_{k=1}^{N} \nu_{k}!  } \left[ \pm  \frac{\lambda_{0}}{\lambda_{T}}            \right]^{\sum_{k=1}^{N}\nu_{k}}  }\,
\mc{P}_{B/F}^{R}(-W)\,\e{\beta W},
\label{Crooks}
\end{equation}
\end{widetext}
where $\lambda_{T}$ is the thermal wave length
\begin{equation}
\lambda_{T}=\sqrt{\frac{2\pi \beta \hbar^{2}}{M}}.
\end{equation}
From Eq.~(\ref{Crooks}) we obtain the difference of the free energy (the Jarzynski Equality \cite{Jarzynski1997a}) of the quantum gas by utilizing the Crooks Fluctuation Theorem \cite{Crooks1998}
\begin{equation}
\begin{split}
&\e{-\beta \Delta F_{B/F}}=\\
&\frac{\sum_{(1^{\nu_{1}},2^{\nu_{2}},\cdots, N^{\nu_{N}} ) } \frac{N!}{ \prod_{k=1}^{N} k^{3\nu_{k}/2}  \prod_{k=1}^{N} \nu_{k}!  } \left[ \pm  \frac{\lambda_{\tau}}{\lambda_{T}}            \right]^{\sum_{k=1}^{N}\nu_{k}}    }                 {\sum_{(1^{\nu_{1}},2^{\nu_{2}},\cdots, N^{\nu_{N}} ) } \frac{N!}{\prod_{k=1}^{N} k^{3\nu_{k}/2}  \prod_{k=1}^{N} \nu_{k}!  } \left[ \pm  \frac{\lambda_{0}}{\lambda_{T}}            \right]^{\sum_{k=1}^{N}\nu_{k}}  }.
\end{split}
\end{equation}
Since $\lambda_{T}$ is very small in the limit of high temperature $\beta \to 0$, the second leading term should come from the $(1^{N-2},2^{1},\cdots,N^{0})$ type permutation, thus $\e{-\beta \Delta F}$ can be approximated as
\begin{equation}
\e{-\beta \Delta F_{B/F}} \approx \frac{\lambda_{\tau}^{N} \pm \frac{N(N-1)}{2^{3/2}} \lambda_{\tau}^{N-1} \lambda_{T} }{\lambda_{0}^{N} \pm \frac{N(N-1)}{2^{3/2}} \lambda_{0}^{N-1} \lambda_{T}}.
\end{equation}
Now we fix $\lambda_{0}$ and replace $\lambda_{\tau}$ with $\lambda$. When the system is in the thermodynamic limit, the free energy difference can be evaluated with the Bernoulli approximation,
\begin{equation}
\Delta F_{B/F}(T,\lambda) \approx -\beta^{-1} N \ln \left[ \frac{\lambda \left(1\pm \frac{1}{2^{3/2}} m \lambda_{T} \right)}{\lambda_{0} \left(1\pm \frac{1}{2^{3/2}} m_{0} \lambda_{T} \right)} \right],
\end{equation}
where $m=N/\lambda$ ($m_{0}=N/\lambda_{0}$) is the particle density. In the following we calculate the pressure of this system by utilizing the thermodynamic relation $p=-(\partial F/ \partial \lambda)_{T}$
\begin{equation}
\begin{split}
p_{B/F}&=-\left( \frac{\partial \Delta F_{B/F}}{\partial \lambda} \right)_{T}\\
& \approx \frac{N \beta^{-1}}{\lambda \pm N \lambda_{T}/2^{3/2}} \approx m \beta^{-1} \left( 1\mp \frac{m \lambda_{T}}{2^{3/2}} \right).
\end{split}
\end{equation}
For a $d$-dimensional system, the equation of state of the quantum gas inside a piston can be obtained in a similar way
\begin{equation}
p_{B/F} \approx m \beta^{-1} \left( 1\mp \frac{m \lambda_{T}^{d}}{2^{1+d/2}} \right).
\label{B7}
\end{equation}
This equation of state (\ref{B7}) agrees with the result derived from the grandcanonical ensemble formulation \cite{Huang1987}. The derivation of the equation of state of ideal quantum gases is an evidence supporting the validity of the characteristic function of the work distribution function (\ref{derivation}). In fact, we have also checked that the first three virial coefficients of the quantum gas are exactly the same as those obtained from the grand canonical ensemble formulation in the thermodynamic limit, which further convinces us the validity of Eq.~(\ref{derivation}).

\section{\label{sec:appC}Heuristic analysis of the convergence of Eq.~(\ref{rewritten}) to Eq.~(\ref{asymptotic}) in the high temperature limit}
Firstly we write the expression of $R_{(1^{\nu_{1}},2^{\nu_{2}},\cdots, N^{\nu_{N}} )}(\beta)$
\begin{equation}
\begin{split}
&R_{(1^{\nu_{1}},2^{\nu_{2}},\cdots, N^{\nu_{N}} )}(\beta)\\
&=\frac{\prod_{k=1}^{N}\left[  \ptr{\mathcal{H}_{s}^{\lambda_{0}}} { \e{-k\beta H_{s}^{\lambda_{0}}}} \right]^{\nu_{k}}}{\left[ \ptr{\mathcal{H}_{s}^{\lambda_{0}}}{\e{-\beta H_{s}^{\lambda_{0}}}}\right]^{N}}.
\label{D1}
\end{split}
\end{equation}
A special case for Eq.~(\ref{D1}) is $R_{(1^{N},2^{0},\cdots, N^{0} )}(\beta)=1$. Also, we know that
\begin{equation}
\sum_{k=1}^{N} k\times \nu_{k}=N.
\label{D2}
\end{equation}
As long as the Hamiltonian has a minimum eigenvalue $E_{0}$ (ground state energy), we can introduce a non-negative definite Hamiltonian $\tilde{H}_{s}^{\lambda_{0}}=H_{s}^{\lambda_{0}}-E_{0}$. By using Eq.~(\ref{D2}), Eq.~(\ref{D1}) can be written as
\begin{equation}
\begin{split}
&R_{(1^{\nu_{1}},2^{\nu_{2}},\cdots, N^{\nu_{N}} )}(\beta)\\
&=\frac{\prod_{k=1}^{N}\left[  \ptr{\mathcal{H}_{s}^{\lambda_{0}}} { \e{-k\beta \tilde{H}_{s}^{\lambda_{0}}}} \right]^{\nu_{k}}}{\left[ \ptr{\mathcal{H}_{s}^{\lambda_{0}}}{\e{-\beta \tilde{H}_{s}^{\lambda_{0}}}}\right]^{N}}.
\label{D4}
\end{split}
\end{equation}
For any permutation $(1^{\nu_{1}},2^{\nu_{2}},\cdots, N^{\nu_{N}} ) \neq (1^{N},2^{0},\cdots, N^{0} )$, from Eq.~(\ref{D2}) we can easily get $\sum_{k=1}^{N} \nu_{k}< N$. Since all eigenvalues of $\tilde{H}_{s}^{\lambda_{0}}$ are nonnegative we also have
\begin{equation}
0< \ptr{\mathcal{H}_{s}^{\lambda_{0}}} { \e{-k\beta \tilde{H}_{s}^{\lambda_{0}}}} \leq \ptr{\mathcal{H}_{s}^{\lambda_{0}}}{\e{-\beta \tilde{H}_{s}^{\lambda_{0}}}}.
\end{equation}
Combining these facts, for any permutation $(1^{\nu_{1}},2^{\nu_{2}},\cdots, N^{\nu_{N}} ) \neq (1^{N},2^{0},\cdots, N^{0} )$, we have
\begin{equation}
\begin{split}
0&< R_{(1^{\nu_{1}},2^{\nu_{2}},\cdots, N^{\nu_{N}} )}(\beta)\\
&\le \left[ \ptr{\mathcal{H}_{s}^{\lambda_{0}}}{\e{-\beta \tilde{H}_{s}^{\lambda_{0}}}}\right]^{\sum_{k=1}^{N} \nu_{k}- N}\\
&\le \left[ \ptr{\mathcal{H}_{s}^{\lambda_{0}}}{\e{-\beta \tilde{H}_{s}^{\lambda_{0}}}}\right]^{-1}.
\end{split}
\end{equation}
As long as the system contains infinitely many energy levels, $\left[ \ptr{\mathcal{H}_{s}^{\lambda_{0}}}{\e{-\beta \tilde{H}_{s}^{\lambda_{0}}}}\right]^{-1}$ will approach zero in the high temperature limit $(\beta \to 0)$. Recalling the fact that $R_{(1^{N},2^{0},\cdots, N^{0} )}(\beta)=1$, we finally prove that
\begin{equation}
\lim_{\beta \to 0} R_{(1^{\nu_{1}},2^{\nu_{2}},\cdots, N^{\nu_{N}} )}(\beta) = \delta_{\nu_{1},N}.
\end{equation}
In addition, we can prove that $R_{(1^{\nu_{1}},2^{\nu_{2}},\cdots, N^{\nu_{N}} )}(\beta)$ is a monotonically non-decreasing function with respect to $\beta$ by directly analyzing the sign of $d \ln{R_{(1^{\nu_{1}},2^{\nu_{2}},\cdots, N^{\nu_{N}} )}(\beta)}/d \beta$
\begin{equation}
\begin{split}
\frac{d}{d \beta}\ln&{R_{(1^{\nu_{1}},2^{\nu_{2}},\cdots, N^{\nu_{N}} )}(\beta)} = N\left\langle E(\beta^{-1},\lambda_{0}) \right\rangle \\
 &-\sum_{k=1}^{N} k \nu_{k} \left\langle E(k^{-1}\beta^{-1},\lambda_{0}) \right\rangle.
 \label{D7}
\end{split}
\end{equation}
Here
\begin{equation}
\left\langle E(\beta^{-1},\lambda_{0}) \right\rangle \equiv \frac{\ptr{\mathcal{H}_{s}^{\lambda_{0}}}{H_{s}^{\lambda_{0}}\e{-\beta H_{s}^{\lambda_{0}}}}}{\ptr{\mathcal{H}_{s}^{\lambda_{0}}}{\e{-\beta H_{s}^{\lambda_{0}}}}}.
\end{equation}
From Eq.~(\ref{D2}) and the fact that $\left\langle E(\beta^{-1},\lambda_{0}) \right\rangle$ must be a non-decreasing function of $T$, we conclude that the right hand side of Eq.~(\ref{D7}) is nonnegative.

So far we have shown that in the high temperature limit the only non-vanishing term in Eq.~(\ref{rewritten}) is the term containing $R_{(1^{N},2^{0},\cdots, N^{0} )}(\beta)$. We can further prove that $|G_{(1^{\nu_{1}},2^{\nu_{2}},\cdots, N^{\nu_{N}} )}(\mu)|\le 1$, which is equivalent to $|\lambda_{1}^{k}+\lambda_{2}^{k}+\lambda_{3}^{k}+\cdots | \le |\lambda_{1}|^{k}+|\lambda_{2}|^{k}+|\lambda_{3}|^{k}+\cdots $ ($|\lambda_{i}|\le 1$ and $\sum_{i=1}^{\infty}|\lambda_{i}| \le \infty $) in the representation of the operator ${\mathcal{G}^{\lambda_{0}} (\mu)\e{-\beta \tilde{H}_{s}^{\lambda_{0}}}}$, with $\{\lambda_{i}\}$ being the complete set of eigenvalues. It seems that we prove the convergence of Eq.~(\ref{rewritten}) to Eq.~(\ref{asymptotic}) since the dominant term in the numerator in the high temperature limit is the term containing $G_{(1^{N},2^{0},\cdots, N^{0} )}(\mu)$. There is, however, one problem in the above derivation. In the high temperature limit, $G_{(1^{\nu_{1}},2^{\nu_{2}},\cdots, N^{\nu_{N}} )}(\mu)$ is probably zero for any nonzero value of $\mu$, from which we can extract no information, despite we know the convergence relation in the sense of absolute value. For example, in the limit $\beta \to 0$, the function $g(\mu)=(1+\alpha \mu/\beta)^{-1/2}$ is trivially zero for any nonzero $\mu$. Thus, we suggest to use $\tilde{G}_{(1^{\nu_{1}},2^{\nu_{2}},\cdots, N^{\nu_{N}} )}(x) \equiv G_{(1^{\nu_{1}},2^{\nu_{2}},\cdots, N^{\nu_{N}} )}(\beta x)$ as a function to demonstrate the convergence of work distributions of three kinds of particles, because it is probably nonvanishing for almost any value of $x$ in the high temperature limit, and this is the reason why we proportionally enlarge the range of the horizontal axis in all the diagrams as the temperature increases.

%

\end{document}